\documentclass[reprint, onecolumn, secnumarabic, amssymb, nobibnotes, aps, prfluids]{revtex4-2}
\usepackage{graphicx}
\usepackage{epstopdf, epsfig}
\usepackage{amsmath}

\usepackage{multirow}
\usepackage[dvipsnames]{xcolor}
\usepackage{cancel}
\usepackage{lineno}
\usepackage{subcaption}
\usepackage{caption}
\usepackage{url}

\providecommand\bnabla{\boldsymbol{\nabla}}
\providecommand\bcdot{\boldsymbol{\cdot}}
\newcommand{\diff}{\mathrm{d}}

\newcommand{\CORR}[1]{{\color{black}{#1}}}
\newcommand{\R}{\mathcal{R}}

\begin{document}
\title{State estimation of Rayleigh-Bénard convection with reduced-order models}

\begin{abstract}
	In this work, we develop a state estimation framework for two-dimensional Rayleigh-Bénard (RB) convection that combines a stable Galerkin reduced-order model (ROM) with an extended Kalman filter (EKF). The ROM, constructed from controllability modes of the linearized Boussinesq equations, provides the nonlinear dynamical model for the filter prediction step. Direct numerical simulations (DNS) are used to generate synthetic measurements for data assimilation. We assess filter performance across periodic, quasiperiodic, and chaotic regimes, demonstrating that the filter tracks the most energetic modes with high fidelity and achieves time-averaged reconstruction errors below \CORR{$16\%$} for velocity and \CORR{$19\%$} for temperature. \CORR{We apply the ROM-based EKF to a hybrid simulation scenario where the system state is assimilated from velocity measurements}. It is shown that velocity observations alone suffice to reconstruct the state, including the temperature field. Finally, we exploit the Kalman gain matrix to develop a greedy sensor placement strategy that progressively removes the least informative sensors. The algorithm reveals a clear hierarchy among sensor types and can be used to derive skeletal observation configurations. It also provides guidance on which measurement variables and spatial locations are most informative for state correction. The present framework is general, and may be applied to other quadratic Galerkin ROMs for state estimation.
\end{abstract}
\author{Enrique Flores-Montoya}
\email{efloresm.ca@gmail.com}
\affiliation{Aerothermals department, ITP Aero, 28108 Alcobendas, Spain}
\author{Andr\'e F. C. da Silva}
\email{andref@ita.br}
\affiliation{Divis\~ao de Engenharia Aeroespacial, Instituto Tecnol\'ogico de Aeron\'autica, S\~ao Jos\'e dos Campos, SP, Brazil}
\author{Andr\'e V. G. Cavalieri}
\email{andre@ita.br}
\affiliation{Divis\~ao de Engenharia Aeroespacial, Instituto Tecnol\'ogico de Aeron\'autica, S\~ao Jos\'e dos Campos, SP, Brazil}
\date{\today}
\maketitle
\section{Introduction}
\label{sec:intro}

Closed-loop flow control has been a long-standing goal in fluid mechanics, with applications ranging from drag reduction and noise suppression to thermal management and mixing enhancement~\cite{brunton2015closed, kim2007linear, bewley2001flow}. Active feedback control requires knowledge of the system state, which is generally unavailable through direct measurement in fluid systems since the state corresponds to spatially distributed fields~\cite{hoepffner2005state}. Consequently, the closed-loop control problem is decomposed into two subproblems: state feedback and state estimation. The latter involves implementing state estimators that reconstruct the system state from either sparse sensor measurements (e.g., pointwise velocity and pressure probes)~\cite{gerhard2003model, ahuja2010feedback} or integral quantities such as aerodynamic forces~\cite{aleksic2010robust}.

The most direct approach to flow state estimation uses direct numerical simulations (DNS), applying data assimilation techniques to the discretized Navier-Stokes equations. 
A significant body of work has focused on the linearized Navier-Stokes equations, which admit rigorous application of classical linear estimation theory. This framework has been used to control transition to turbulence in wall-bounded flows, where linear dynamics provide a useful approximation during the early stages of transition~\cite{bagheri2009input,jovanovic2005componentwise}.
\citet{hogberg2003linear} used Kalman filter-based estimators from wall measurements for transition control in plane channel flow, achieving substantial increases in transition thresholds.
\citet{hoepffner2005state} and~\citet{chevalier2006state} applied Kalman filter (KF) estimators to wall-bounded flows in a two-part study, covering perturbed laminar and fully turbulent regimes, respectively.
The high-dimensionality of flow systems, where the state dimension can reach $\mathcal{O}(10^6)$, makes the application of standard KF implementations computationally demanding~\cite{brunton2015closed, da2019enkf, da2018ensemble}. 
Ensemble methods avoid explicit covariance propagation, making them tractable for high-dimensional systems.
\citet{colburn2011state} applied ensemble Kalman filters (EnKF) to turbulent channel flow, extending earlier work on standard and extended Kalman filters (EKF) for state estimation in laminar and turbulent channel flows~\cite{hoepffner2005state,chevalier2006state}. 
\citet{da2018ensemble} developed EnKF-based estimators for the 2D flow past a cylinder and an airfoil at high angle of attack and low Reynolds number.

The high computational cost of full-order models has motivated the use of reduced-order models (ROMs) in closed-loop control applications~\cite{rowley2017model,brunton2022data, brunton2015closed}. In Galerkin ROMs, the governing partial differential equations are projected onto a set of spatial modes -- typically obtained from proper orthogonal decomposition (POD) of simulation snapshots -- reducing the problem to a system of ordinary differential equations for the modal amplitudes~\cite{rowley2017model}. This approach comes with some loss in model accuracy but achieves a large reduction of the problem dimension, from $\mathcal{O}(10^6)$ unknowns to $\mathcal{O}(10)$--$\mathcal{O}(10^2)$. At this scale, classical control and estimation methods become tractable, with several successful applications reported in the literature~\cite{gerhard2003model, aleksic2010robust,ahuja2010feedback, flinois2016feedback}.  Moreover, the low computational cost of the ROMs enables real-time integration of the dynamical system, which is needed for practical implementation of control systems in experimental settings.

These control applications share a common requirement: the feedback law is expressed in terms of modal amplitudes, which must be estimated from physical measurements in real time. Full flow field measurements are generally not available, and sensing is typically restricted to sparse observations with limited spatial coverage. State observers that combine the ROM dynamics with such sparse measurements are therefore required.
\citet{gerhard2003model} reconstructed a cylinder wake state from a single hot-wire signal using a dynamic observer that runs the Galerkin model in parallel with the flow, correcting the estimated modal coefficients based on the sensor discrepancy.
\citet{rowley2005model} constructed dynamic observers for cavity flow oscillations using a POD-Galerkin ROM. Comparing the flow state reconstructions from a single wall-pressure sensor, they found that the KF outperforms the linear stochastic estimation method, especially for noisy sensor measurements.
\citet{aleksic2010robust} used an EKF to estimate the mean-field modal amplitudes from the spatially averaged base pressure measured at the bluff body stern.
\citet{ahuja2010feedback} extended the approximate balanced truncation method to unstable linear systems and used the resulting ROM to design a KF that reconstructs the flow state from two near-wake velocity measurements, enabling stabilization of an unstable steady state of flow past a flat plate.
\citet{john2010pod} combined a second-order KF with a POD-Galerkin ROM for the incompressible Navier–Stokes equations with a passive scalar field to estimate velocity and concentration from sparse point measurements.
\citet{kikuchi2015assessment} combined a POD-Galerkin ROM with ensemble-based data assimilation methods to estimate the unsteady flow past a cylinder from sparse velocity observations in the wake. They compared an EnKF and a particle filter (PF), finding that the PF produced lower error when the state distribution was non-Gaussian. 
\citet{sommer2023estimating} combined a POD-Galerkin ROM with an EKF to reconstruct the velocity and pressure fields inside a centrifugal pump from velocity measurements at a small number of locations. A greedy algorithm is used to select measurement positions that minimise the condition number of the observability matrix, and the EKF is shown to provide stable estimates over long times even when the underlying ROM itself is unstable.

Despite these remarkable examples, traditional POD-based Galerkin ROMs suffer from a major limitation~\cite{cavalieri2022reduced}. They require closure models that introduce additional dissipation into the system to prevent numerical instability~\cite{ahmed2021closures, soucasse2019proper}. 
An alternative methodology has recently emerged that addresses this limitation by constructing modal bases from the linearized equations rather than from snapshot data. \citet{cavalieri2022reduced} developed Galerkin ROMs for three-dimensional plane Couette flow using orthogonal bases obtained from the controllability Gramian of the linearized Navier-Stokes equations. 
The resulting ROMs remain numerically stable without closure models and were used in~\cite{mccormack2024multi} to compute multi-scale invariant solutions and stable orbits of the system. This framework was extended to a 2D mixing layer in ~\cite{cavalieri2023non}. In a recent work,~\citet{maia2025turbulence} demonstrated that these stable ROMs could be used to design control inputs that laminarize turbulent plane Couette flow.

Beyond flow control, data assimilation techniques are also employed in hybrid simulations. These approaches combine experimental measurements with numerical simulations, typically DNS, to reconstruct flow quantities that are difficult to measure directly or to overcome the temporal and spatial resolution limitations of experimental techniques~\cite{nisugi2004fundamental, suzuki2009unsteady1, suzuki2009unsteady2, suzuki2012reduced, hayase2015review}.
Rayleigh-Bénard (RB) convection provides a particularly well-suited testbed for such approaches, owing to its rich dynamical behavior, well-characterized bifurcation structure, and relevance for technological applications such as crystal growth~\cite{muller1988convection, alloui2018control, gunzburger2002controlling}. 
In this setting,~\citet{bauer2022assimilation} assimilated tomographic particle image velocimetry (PIV) measurements into a DNS via nudging, reconstructing fully resolved flow fields from measurements.
\citet{cornick2009state} applied the local ensemble transform Kalman filter (LETKF) to state estimation in chaotic RB convection, reconstructing the temperature and velocity fields from shadowgraph images in both numerical and laboratory experiments.
\citet{farhat2020data} established rigorous synchronization conditions for data assimilation from thermal measurements alone, proving that temperature observations at sufficiently fine spatial resolution guarantee convergence to the true state. 
\citet{agasthya2022reconstructing} and~\citet{clarcdileoni2023reconstructing} reconstructed velocity fields from temperature-only measurements in turbulent 2D RB convection using nudging and physics-informed neural networks, respectively.

In our companion paper~\citep{flores2025galerkin}, we developed stable Galerkin ROMs for 2D RB convection using controllability modes of the linearized Boussinesq equations and validated them against DNS in terms of mean profiles, heat flux, flow structures, and dynamical regimes. Motivated by the need for computationally efficient state estimators capable of operating across different dynamical regimes, the present work demonstrates a practical application of these ROMs: their use as dynamical models within an EKF for real-time state estimation. Using DNS as synthetic measurements, the filter is evaluated across periodic, quasiperiodic, and chaotic regimes to systematically compare its performance and robustness (\S\ref{sec:coarse_grid}). Then, we consider a hybrid simulation scenario where the state is assimilated from \CORR{velocity measurements} (\S\ref{sec:velocity_only}). The objective is to show that the present state estimation framework can be used to infer the temperature field without direct thermal observations. In \S\ref{sec:sensor_placement}, we develop a greedy sensor placement strategy based on the Kalman gain matrix that progressively removes the least informative sensors and can be used to derive skeletal observation strategies, minimizing the number of sensors required for state estimation. Finally, in \S\ref{sec:estimated_cov}, we estimate the noise covariances from DNS data and compare their performance with the isotropic, uncorrelated noise covariance matrices used throughout the rest of the study.

\CORR{In this work, we consider low to moderate Rayleigh numbers ($Ra\lesssim2\times10^5$), which is the range where our Galerkin ROMs have been validated in terms of mean profiles, heat flux scaling, spectra and dynamical regime~\citep{flores2025galerkin}. Although moderate, this range is relevant for a number of technological applications where thermal convection takes place in small enclosures or at moderate temperature differences, such as natural-convection cooling of electronics~\cite{alexandersen2014topology,saglietti2018topology}, chemical vapour deposition~\cite{jensen1991flow}, crystal growth~\cite{langlois1985buoyancy,lappa2005thermal}, convective PCR~\cite{muddu2011chaotically,priye2013microscale} and insulating window cavities~\cite{basok2022evaluation}. 
Furthermore, active feedback control to suppress convection in RB, which requires estimating the state of the system from measurements, has so far been demonstrated, to the best of our knowledge, only at similar or lower Rayleigh numbers ($Ra\lesssim3\times10^4$)~\cite{tang1993stabilization,tang1998experiments,howle1997active,remillieux2007suppression,or2001robust,or2003active,beintema2020controlling,jiren2025comprehensive}.}

The remainder of this paper is organized as follows. \S\ref{sec:theory} presents the governing equations, the Galerkin ROM formulation, and the EKF algorithm. \S\ref{sec:numerics} describes the numerical methods for ROM generation, DNS, and filter implementation. Results are presented in \S\ref{sec:results} and conclusions are given in \S\ref{sec:conclusions}.

\section{Theoretical background}\label{sec:theory}
We consider two-dimensional Rayleigh-Bénard convection in a domain periodic in the horizontal direction $x$ and bounded by no-slip walls at $y=0$ and $y=1$. The non-dimensional governing equations for a Boussinesq fluid are the continuity, momentum, and energy conservation equations:
\begin{eqnarray}\label{eq:cont}
\bnabla\bcdot\boldsymbol{u} = 0,\\
\label{eq:mom}
\frac{\partial \boldsymbol{u}}{\partial t} + \boldsymbol{u}\bcdot\bnabla\boldsymbol{u} = -\nabla p + Pr\,\theta\,\boldsymbol{e}_y  + \frac{Pr}{\sqrt{Ra}}\bnabla^2\boldsymbol{u},\\
\label{eq:ener}
\frac{\partial \theta}{\partial t} + \boldsymbol{u}\bcdot\nabla\theta = \frac{1}{\sqrt{Ra}}\bnabla^2\theta,
\end{eqnarray}
where $\boldsymbol{u} = \begin{bmatrix}u & v\end{bmatrix}^T$ and $\theta$ are the dimensionless velocity and temperature fields, respectively, and $p$ is the driven pressure. Bold symbols are used to denote vector variables and $\boldsymbol{e}_y$ is the unit vector in the $y$ direction.
The Prandtl number $Pr = \nu/\kappa$ represents the ratio between the viscous and thermal diffusion coefficients. 
The ratio between buoyancy forces and viscous dissipation is quantified through the Rayleigh number, $Ra = \sigma g h^3 \Delta T / (\nu\kappa)$, where $\sigma$ is the fluid thermal expansion coefficient, $g$ is the gravitational acceleration, $h$ is the dimensional distance between the walls and $\Delta T$ is the imposed temperature jump. In eqs.~\eqref{eq:cont}--\eqref{eq:ener}, velocity and temperature have been normalized using $\kappa \sqrt{Ra}/h$ and $\Delta T$, respectively.

\subsection{Galerkin reduced-order model}
In the companion paper~\cite{flores2025galerkin}, coupled-basis ROMs are shown to outperform uncoupled-basis models in terms of total heat flux and turbulence statistics. For this reason, in the present work we adopt a coupled-basis approach following~\cite{podvin2001low, podvin2012proper, soucasse2019proper, flores2025galerkin}. For that, we define an extended state vector that groups the velocity and temperature fields:
\begin{equation}\label{eq:extended_state}
    \mathcal{X} = \begin{bmatrix}
        u & v & \theta
    \end{bmatrix}^T.
\end{equation}
The extended state is decomposed into a baseline steady solution, $\theta_0(y)$, and a time-dependent perturbation expressed as a sum of orthonormal modes:
\begin{equation}\label{eq:modal_decomposition}
    \mathcal{X}(t, \boldsymbol{x}) = \begin{bmatrix}
        0 & 0 & \theta_0(y)
    \end{bmatrix}^T + \sum_{j=1}^{n} c_j(t)\,\boldsymbol{\chi}_j(\boldsymbol{x}),
\end{equation}
where $\theta_0(y) = 1 - y$, is the conductive equilibrium temperature profile and $\boldsymbol{\chi}_j(\boldsymbol{x})$ are coupled modes containing velocity and temperature components. Each mode $\boldsymbol{\chi}_j$ can be written as:
\begin{equation}
    \boldsymbol{\chi}_j(\boldsymbol{x}) = \begin{bmatrix}
        \mathcal{U}_j(\boldsymbol{x}) &
        \mathcal{V}_j(\boldsymbol{x}) &
        \mathcal{T}_j(\boldsymbol{x})
    \end{bmatrix}^T,
\end{equation}
where $\mathcal{U}_j$, $\mathcal{V}_j$ and $\mathcal{T}_j$ are the horizontal velocity, vertical velocity and temperature components of mode $j$, respectively. From this decomposition, the physical fields are reconstructed as:
\begin{equation}\label{eq:field_reconstruction}
    \boldsymbol{u}(t, \boldsymbol{x}) = \sum_{j=1}^{n} c_j(t)\,\boldsymbol{\mathcal{V}}_j(\boldsymbol{x}), \qquad
    \theta(t, \boldsymbol{x}) = \theta_0(y) + \sum_{j=1}^{n} c_j(t)\,\mathcal{T}_j(\boldsymbol{x}),
\end{equation}
where $\boldsymbol{\mathcal{V}}_j = \begin{bmatrix}\mathcal{U}_j & \mathcal{V}_j
\end{bmatrix}^T$ groups the velocity components of mode $j$. 
For the coupled formulation, a weighted inner product is introduced to balance the energy contributions from velocity and temperature fields:
\begin{equation}\label{eq:inner_product}
    \langle\mathcal{X}_i,\,\mathcal{X}_j\rangle_c = \frac{1}{L_x 
    L_y}\int_0^{L_x}\int_{0}^{L_y}\left(u_i u_j + v_i v_j + 
    \gamma^2\theta_i\theta_j\right)\diff y\,\diff x,
\end{equation}
where $\gamma^2$ is a weighting factor representing the ratio between the average kinetic and thermal perturbation energies. \CORR{In eq.~\eqref{eq:inner_product}, $L_x$ and $L_y$ denote the horizontal and vertical dimensions of the fluid domain, respectively}.
 Following \citet{flores2025galerkin}, the weighting factor is set to $\gamma^2 = 1.24$, a value determined from a least-squares fit of DNS data.
The coupled modes $\boldsymbol{\chi}_j$ form an orthonormal basis under this inner product:
\begin{equation}\label{eq:orthogonality_cond}
    \langle\boldsymbol{\chi}_i,\,\boldsymbol{\chi}_j\rangle_c = \delta_{ij}.
\end{equation}
Substituting the modal decomposition from eq.~\eqref{eq:modal_decomposition} into the governing equations eqs.~\eqref{eq:mom}--\eqref{eq:ener} and taking the inner product with $\boldsymbol{\chi}_i$ yields the ROM ordinary differential equation:
\begin{equation}\label{eq:ROM}
    \frac{\diff c_i}{\diff t} = Pr\left(\mathcal{F}^0_i + \sum_j \mathcal{F}^1_{ij}\,c_j\right) + \frac{Pr}{\sqrt{Ra}}\sum_j \mathcal{D}^{\mathcal{V}}_{ij}\,c_j + \frac{1}{\sqrt{Ra}}\sum_j \mathcal{D}^{\mathcal{T}}_{ij}\,c_j - \sum_j \mathcal{L}_{ij}\,c_j - \sum_{j,k} \mathcal{N}_{ijk}\,c_j\,c_k,
\end{equation}
where the projected operators are defined as:
\begin{align}
    \mathcal{F}^0_i &= \langle\boldsymbol{\chi}_i,\,\theta_0(y)\,\boldsymbol{e}_v\rangle_c, \\
    \mathcal{F}^1_{ij} &= \langle\boldsymbol{\chi}_i,\,\mathcal{T}_j\,\boldsymbol{e}_v\rangle_c,\\
    \label{eq:vel_diff}\mathcal{D}^{\mathcal{V}}_{ij} &= \langle\boldsymbol{\mathcal{V}}_i,\,\nabla^2\boldsymbol{\mathcal{V}}_j\rangle,\\
    \label{eq:tem_diff}\mathcal{D}^{\mathcal{T}}_{ij} &= \gamma^2\langle\mathcal{T}_i,\,\nabla^2\mathcal{T}_j\rangle,\\
    \mathcal{L}_{ij} &= \langle\boldsymbol{\chi}_i,\,(\boldsymbol{\mathcal{V}}_j\cdot\nabla\theta_0)\,\boldsymbol{e}_\theta\rangle_c,\\
    \mathcal{N}_{ijk} &= \langle\boldsymbol{\chi}_i,\,\boldsymbol{\mathcal{V}}_k\cdot\nabla\boldsymbol{\chi}_j\rangle_c,
\end{align}
with $\boldsymbol{e}_v = \begin{bmatrix} 0 & 1 & 0 \end{bmatrix}^T$ and $\boldsymbol{e}_\theta = \begin{bmatrix} 0 & 0 & 1 \end{bmatrix}^T$ being unit vectors in the wall-normal velocity and temperature components, respectively. In eqs.~\eqref{eq:vel_diff}--\eqref{eq:tem_diff}, the inner product $\langle \boldsymbol{\phi},\,\boldsymbol{\psi}\rangle$ (without subscript $c$) is defined as:
\begin{equation}
	\langle \boldsymbol{\phi},\,\boldsymbol{\psi}\rangle = \frac{1}{L_xL_y}\int_0^{L_x}\int_0^{L_y}\boldsymbol{\phi}\cdot\boldsymbol{\psi}\,\text{d}y\,\text{d}x.
\end{equation}
The ROM differential equation eq.~\eqref{eq:ROM} can be written in compact form as:
\begin{equation}\label{eq:ROM_compact}
    \frac{\diff \boldsymbol{c}}{\diff t} = \boldsymbol{f}(\boldsymbol{c}; Ra, Pr),
\end{equation}
where $\boldsymbol{c} = \begin{bmatrix} c_1, & c_2,& \ldots, &c_n\end{bmatrix}^T$ is the vector of modal amplitudes and $\boldsymbol{f}$ represents the right-hand side of eq.~\eqref{eq:ROM}. To build a Galerkin ROM it is necessary to generate a modal basis satisfying the orthogonality condition expressed in eq.~\eqref{eq:orthogonality_cond}. Coupled orthonormal modal bases can be obtained as the eigenfunctions of the controllability Gramian of the linearized RB equations. The procedure to obtain compliant bases and the derivation details are thoroughly described in~\cite{flores2025galerkin}.
The quadratic nonlinearity of eq.~\eqref{eq:ROM_compact} yields an analytical expression for the Jacobian of $\boldsymbol{f}$ with respect to $\boldsymbol{c}$, which is exploited in \S\ref{sec:Kalman} for the linearization required by the EKF prediction step and avoids the need for finite-difference approximations.

\subsection{Extended Kalman filter}\label{sec:Kalman}
The objective is to estimate the modal amplitudes $\boldsymbol{c}(t)$ from sparse measurements of the physical fields obtained from direct numerical simulations. This is formulated as a state estimation problem where the ROM provides the dynamical model and the measurements constitute observations of the system. The ROM dynamics from eq.~\eqref{eq:ROM_compact} are discretized in time to obtain a discrete-time state-space model:
\begin{equation}\label{eq:state_evolution}
    \boldsymbol{c}_{k} = \boldsymbol{\Phi}(\boldsymbol{c}_{k-1}; Ra, Pr, \Delta t_k) + \boldsymbol{w}_{k-1},
\end{equation}
where $\boldsymbol{c}_k$ denotes the state at time $t_k$, $\boldsymbol{\Phi}$ represents the nonlinear state transition obtained by integrating eq.~\eqref{eq:ROM_compact} from $t_{k-1}$ to $t_k$, and \CORR{$\boldsymbol{w}_{k-1} \sim \mathrm{N}(\boldsymbol{0}, Q)$} is the process noise accounting for model uncertainty. \CORR{Here, $\mathrm{N}(\mathbf{0}, Q)$ denotes a zero mean Gaussian noise with covariance matrix $Q$.}

Measurements are obtained at discrete spatial locations (probes) in the DNS domain. Let $\boldsymbol{y}_k \in \mathbb{R}^{m}$ denote the observation vector \CORR{with dimension $m$} at time $t_k$, containing measurements of velocity and temperature at the probe locations. The observation model relates the measurements to the modal amplitudes:
\begin{equation}\label{eq:observation}
    \boldsymbol{y}_k = H\,\boldsymbol{c}_k + \boldsymbol{v}_k,
\end{equation}
where $H \in \mathbb{R}^{m \times n}$ is the observation matrix and \CORR{$\boldsymbol{v}_k \sim \mathrm{N}(\boldsymbol{0}, R)$ is the Gaussian measurement noise with zero mean and covariance $R$}. The uncertainty in the measurements accounts for off-projection errors from the full order system on the reduced order modal basis.
The standard KF formulation treats this noise as white in time, so that $\boldsymbol{v}_k$ is uncorrelated between successive time steps. In the present setup, this does not strictly hold: the off-projection error is affected by the advection of unresolved flow structures past the probes, which evolves smoothly over consecutive snapshots rather than re-randomising at each step.  Accounting for this properly would require a coloured-noise model which is out of the scope of this work. Here, we retain the white noise assumption for simplicity and leave the extension to autoregressive measurement models for future work.

The observation matrix $H$ is constructed from the modal basis evaluated at the probe locations. 
\CORR{In this work, we have restricted the observations to direct velocity and temperature measurements. Despite their high relevance for experimental applications, to constrain the scope of the study we deliberately omitted wall sensors measuring the shear stress or the heat flux. Future works could explore combining gradient-type sensors with direct measurements.}
We consider $n_u$ horizontal velocity, $n_v$ wall-normal velocity and $n_\theta$ temperature probes.
Let $\boldsymbol{x}_\ell$, $\boldsymbol{x}_p$ and $\boldsymbol{x}_q$ denote the position of the $\ell$-th horizontal velocity, $p$-th vertical velocity and $q$-th temperature probes, respectively. The observation matrix has the block structure:
\begin{equation}\label{eq:obs_matrix}
    H = \begin{bmatrix}
        H^u &
        H^v &
        H^\theta
    \end{bmatrix}^T,
\end{equation}
where the sub-matrices are:
\begin{equation}
    H^u_{\ell j} = \mathcal{U}_j(\boldsymbol{x}_\ell), \quad
    H^v_{pj} = \mathcal{V}_j(\boldsymbol{x}_p), \quad
    H^\theta_{qj} = \mathcal{T}_j(\boldsymbol{x}_q).
\end{equation}
The observation vector at each time step is assembled from the DNS fields as:
\begin{equation}\label{eq:meas_vect}
    \boldsymbol{y}_k = [u(\boldsymbol{x}_1, t_k),\,\ldots,\,u(\boldsymbol{x}_\ell, t_k),\,\ldots,\, u(\boldsymbol{x}_{n_u}, t_k),\, \ldots,\,{v}(\boldsymbol{x}_p, t_k),\,\ldots,\, {v}(\boldsymbol{x}_{n_v}, t_k) ,\ldots,\,{\theta}^{\prime}(\boldsymbol{x}_q, t_k) ,\,\ldots ,\,{\theta}^{\prime}(\boldsymbol{x}_{n_\theta}, t_k)]^T,
\end{equation}
where $\theta^{\prime} = \theta - \theta_0$ is the temperature perturbation.

Due to the nonlinear dynamics of the ROM, the EKF is employed. The EKF linearizes the state transition about the current estimate and proceeds in two steps at each time instant: prediction and update.

\subsubsection{Prediction step}
Given the posterior estimate $\hat{\boldsymbol{c}}_{k-1}$ and covariance $P_{k-1}$ at time $t_{k-1}$, the predicted state at time $t_k$ is obtained by integrating the ROM:
\begin{equation}\label{eq:prediction_state}
    \hat{\boldsymbol{c}}_{k}^{-} = \boldsymbol{\Phi}(\hat{\boldsymbol{c}}_{k-1}; Ra, Pr, \Delta t_k).
\end{equation}
The predicted covariance is computed using the linearized dynamics:
\begin{equation}\label{eq:prediction_cov}
    P_k^{-} = F_k\,P_{k-1}\,F_k^T + Q,
\end{equation}
where $Q$ is the model error covariance matrix, which accounts for the uncertainty introduced at each prediction step by the unresolved dynamics in the ROM, and $F_k$ is the state transition matrix obtained from the Jacobian of the ROM:
\begin{equation}\label{eq:Jacobian}
    F_k = I + J(\hat{\boldsymbol{c}}_{k-1})\,\Delta t_k, \qquad J_{ij} = \frac{\partial f_i}{\partial c_j}.
\end{equation}
\CORR{In eqs.~\eqref{eq:prediction_state} and~\eqref{eq:prediction_cov}, superscript $-$ denotes an \emph{a priori} estimate of the Kalman filter i.e., the predicted state and covariance before the measurement update is applied.} The Jacobian of the ROM dynamics is:
\begin{equation}
    J = Pr\,\mathcal{F}^1 + \frac{Pr}{\sqrt{Ra}}\mathcal{D}^{\mathcal{V}} + \frac{1}{\sqrt{Ra}}\mathcal{D}^{\mathcal{T}} - \mathcal{L} - \mathcal{N}\boldsymbol{c} - \mathcal{N}^T\boldsymbol{c},
\end{equation}
where $(\mathcal{N}\boldsymbol{c})_{ij} = \sum_k \mathcal{N}_{ijk}\,c_k$ and $(\mathcal{N}^T\boldsymbol{c})_{ij} = \sum_k \mathcal{N}_{ikj}\,c_k$.  Here the ROM structure presents an advantage, as the Jacobian maintains moderate dimensions and is known analytically.

\subsubsection{Update step}
When a measurement, $\boldsymbol{y}_k$, becomes available, the innovation is computed:
\begin{equation}\label{eq:innovation}
    \boldsymbol{z}_k = \boldsymbol{y}_k - H\,\hat{\boldsymbol{c}}_k^{-}.
\end{equation}
The innovation covariance is:
\begin{equation}\label{eq:innovation_cov}
    S_k = H\,P_k^{-}\,H^T + R,
\end{equation}
where $R$ is the measurement covariance matrix. The Kalman gain determines the optimal weighting between the model prediction and the measurement:
\begin{equation}\label{eq:Kalman_gain}
    K_k = P_k^{-}\,H^T\,S_k^{-1}.
\end{equation}
The posterior state estimate incorporates the measurement:
\begin{equation}\label{eq:update_state}
    \hat{\boldsymbol{c}}_k = \hat{\boldsymbol{c}}_k^{-} + K_k\,\boldsymbol{z}_k.
\end{equation}
The posterior covariance is updated using the Joseph form for numerical stability:
\begin{equation}\label{eq:update_cov}
    P_k = (I - K_k H)\,P_k^{-}\,(I - K_k H)^T + K_k\,R\,K_k^T.
\end{equation}
The performance of the EKF depends on the specification of the process noise covariance $Q$ and measurement noise covariance $R$. The process noise $Q$ accounts for model errors including unresolved dynamics arising from the truncation of the ROM. The measurement noise $R$ represents sensor noise and uncertainty in the observations.

The relative magnitude between $Q$ and $R$ controls the balance between trusting the model prediction and trusting the measurements. A larger $Q$ relative to $R$ results in higher Kalman gains, giving more weight to the measurements. The initial state covariance $P_0$ is typically set to a diagonal matrix reflecting uncertainty in the initial state estimate.

\subsubsection{Error metrics}
The quality of the state estimation is quantified via the following instantaneous error metrics:
\begin{enumerate}
    \item The \CORR{normalized filter} error, $e_\mathbf{c} = {\|\hat{\boldsymbol{c}} -\boldsymbol{c}\|}/{\|\boldsymbol{c}\|}$, where $\boldsymbol{c}$ is the modal amplitude vector obtained by projecting the DNS fields onto the modal basis and $\hat{\boldsymbol{c}}$ is the filter estimate.
    \item The normalized velocity error:
\begin{equation}\label{eq:error_vel}
    e_\mathbf{u} = \dfrac{\left(\int_\Omega (\boldsymbol{u}- \hat{\boldsymbol{u}})^2 d\Omega\right)^{1/2}}{\left(\int_\Omega \boldsymbol{u}^2 d\Omega\right)^{1/2}},
\end{equation}
where $\boldsymbol{u}$ is the DNS velocity field and $\hat{\boldsymbol{u}}$ is the velocity field reconstructed from the filter estimates:
\begin{equation}
    \hat{\boldsymbol{u}} (t, \boldsymbol{x}) = \sum_{j=1}^n \hat{c}_{j}(t) \boldsymbol{\mathcal{V}}_j(\boldsymbol{x}).
\end{equation}
In eq.~\eqref{eq:error_vel}, $\int_\Omega\cdot \,\text{d}\Omega$ denotes integration over the fluid domain.
    \item The normalized temperature error:
\begin{equation}\label{eq:error_temp}
    \CORR{e_\theta = \dfrac{\left(\int_\Omega (\theta - \hat{\theta})^2 d\Omega\right)^{1/2}}{\left(\int_\Omega {\theta^\prime}^2 d\Omega\right)^{1/2}},}
\end{equation}
where ${\theta}$ is the DNS temperature field and $\hat{\theta}$ is the temperature field reconstructed from the filter estimates:
\begin{equation}
    \CORR{\hat{\theta} (t, \boldsymbol{x}) = \theta_0+  \sum_{j=1}^n \hat{c}_{j}(t) {\mathcal{T}}_j(\boldsymbol{x})}.
\end{equation}
\end{enumerate}
It is important to note that the velocity and temperature errors, $e_\mathbf{u}$ and $e_\theta$, are bounded from below by the ROM truncation error. The projection of the DNS fields onto a finite number of coherent spatial structures necessarily discards information contained in the unresolved modes. Consequently, even a perfect estimate of the modal amplitudes ($\hat{\boldsymbol{c}} = \boldsymbol{c}$) would still yield nonzero values of $e_\mathbf{u}$ and $e_\theta$.
For each of the above-described error metrics, the total error is computed as:
\begin{equation}
    E = \frac{1}{T}\int_0^T e(t^\prime)\,\mathrm{d}t^\prime 
\end{equation}
using the trapezoidal rule to integrate discrete signals sampled at $t_k$. \CORR{Since the error metrics are defined using the instantaneous discrepancy between the estimate and the true state, they penalize both the mean and fluctuating error, functioning similarly to a normalized root-mean-square error (RMSE).}

\section{Numerical methods}\label{sec:numerics}
A two-dimensional domain with $L_x = 2$ \CORR{and $L_y=1$} is considered. This aspect ratio allows perturbations with the critical wavenumber, $k_c$, to develop; resulting in a critical Rayleigh number of $Ra_c = 1707.8$. The procedure for generating coupled orthonormal modal bases follows~\citet{flores2025galerkin}. 
Modes are obtained as the eigenfunctions of the Gramian of the linearized Rayleigh-Bénard equations. The periodicity of the domain allows decomposing the flow into Fourier modes in the $x$ direction, so that the problem reduces to a set of independent one-dimensional eigenvalue problems for each wavenumber $k_x = j\alpha$, where $\alpha = 2\pi/L_x$ and $j = 0, 1, \ldots, n_\alpha - 1$. The $y$ direction is discretized using $n_y = 64$ Chebyshev \CORR{modes}. For each wavenumber, a Sylvester equation is solved using LAPACK routines~\cite{anderson1999lapack} to obtain the Gramian. The number of modes retained per wavenumber is $n_\beta$, yielding a total of $n = n_\alpha \times n_\beta$ modes, i.e., $n$ degrees of freedom (DoF) in the ROM. The coupled bases are generated with $Ra = 1$, $Pr = 1$ and the energy-weighting factor $\gamma^2 = 1.24$, determined from a least-squares fit of DNS data as described in~\citet{flores2025galerkin}. The ROM coefficients $\mathcal{F}^0_i$, $\mathcal{F}^1_{ij}$, $\mathcal{D}^{\mathcal{V}}_{ij}$, $\mathcal{D}^{\mathcal{T}}_{ij}$, $\mathcal{L}_{ij}$ and $\mathcal{N}_{ijk}$ are computed by projecting the governing equations onto the two-dimensional modes. To prevent aliasing in the nonlinear terms, the number of grid points in $x$ is set to $4(n_\alpha - 1) + 2$. Table~\ref{tab:mod_comp} lists the parameters of the ROMs used in this work. 

\begin{table}
	\centering
	\nolinenumbers
	\CORR{
	\begin{tabular}{cccccc}
	Nomenclature & $n_{\alpha}$ & $n_{\beta}$ & $n$ & $n_x$ & $n_y$     \\ \hline
			 C96           & 6            & 16     & 96  & 22    & 64   \\
			 C192          & 8            & 24         & 192 & 30    & 64      \\\hline
		\end{tabular}
		\caption{Modal composition, generation parameters and mesh resolution of the different models evaluated. \CORR{All coupled modal bases are generated with $Ra=1$, $Pr=1$ and $\gamma^2=1.24$.}}
		\label{tab:mod_comp}
	}
\end{table}

DNS are performed using the Dedalus spectral solver~\cite{burns2020dedalus}. The domain is discretized with $n_x = 128$ Fourier modes in $x$ and $n_y = 64$ Chebyshev \CORR{modes} in $y$, with a dealiasing factor of $3/2$ in both directions.
 The time integration of eqs.~\eqref{eq:cont}--\eqref{eq:ener} is performed using a second-order semi-implicit backwards differentiation formula (SBDF2)~\cite{burns2020dedalus}. 
 Simulations are initialized from a randomly perturbed conductive state $\theta_0 = 1 - y$ and integrated over a total time $T_{\mathrm{sim}}=1500\,Pr^{-1/2}$, discarding the first third of each run to remove initial transients. The remaining interval, spanning for a simulation time $T = 1000\,Pr^{-1/2}\approx316$, is used for filter assimilation. Flow snapshots are saved every $\Delta t = 3/(5\sqrt{Pr})\approx0.19$ time units, yielding $N_s = 1666$ snapshots per simulation. The random seed $\eta$ controlling the initial perturbation is varied to generate independent realizations of the same flow configuration. Table~\ref{tab:dns_params} lists the DNS realizations and their parameters.

\begin{table}[h]
	\centering
	\nolinenumbers
	\CORR{
\begin{tabular}{cccccc}
	$\R = Ra/Ra_c$       & $\quad Ra\quad$             & $\quad\eta\quad$ & $\quad\langle Nu\rangle\quad$ & $\quad Re_\tau\quad$ & $\quad y_w^+\quad$ \\ \hline
	40  &  68\,310 &                 1                & 4.15                          & 11.2                 & 0.015              \\
	80  & 136\,621 &                1                & 4.84                          & 15.0                 & 0.020               \\
	120 &204\,931 &                1                & 5.13                          & 17.2                 & 0.023              \\
	120& 204\,931&                 2                & 5.15                          & 17.2                 & 0.023              \\
	120& 204\,931&                 3                & 5.13                          & 17.1                 & 0.023              \\
	120& 204\,931&                 4                & 5.13                          & 17.2                 & 0.023        \\
\hline     
\end{tabular}
	\caption{DNS realizations and their parameters. $\eta$ denotes the random seed controlling the initial perturbation. The Nusselt number $\langle Nu\rangle$, friction Reynolds number $Re_\tau$, and wall-normal resolution in wall units $y_w^+$ are defined in~\citet{flores2025galerkin}. \CORR{The Prandtl number is $10$ for all the DNS.}}
	\label{tab:dns_params}
}
\end{table}

Measurements for the EKF are extracted from the DNS fields at discrete probe locations and assembled into the observation vector $\boldsymbol{y}_k$ as described in \S\ref{sec:Kalman}. The DNS solution at each snapshot is projected onto the ROM modal basis to obtain reference coefficients $\boldsymbol{c}_k$. This projection is performed by evaluating the weighted inner product defined in eq.~\eqref{eq:inner_product} between the DNS fields and each mode $\boldsymbol{\chi}_j$ after upsampling the modes to the DNS resolution via zero padding:
\begin{equation}\label{eq:dns_projection}
	c_{i,k} = \langle \mathcal{X}_k^\mathrm{DNS},\,\boldsymbol{\chi}_i \rangle_c, \qquad i = 1,\ldots,n,
\end{equation}
where $\mathcal{X}_k^\mathrm{DNS} = [u_k,\, v_k,\, \theta_k - \theta_0]^T$ is the extended DNS state at time $t_k$. The integrals are computed using the trapezoidal rule in $x$ and Clenshaw--Curtis quadrature in $y$.

The EKF is initialized from $\hat{\boldsymbol{c}}_0 =\hat{\boldsymbol{c}}_0^- = \boldsymbol{0}$ and with an initial covariance $P_0 = \alpha_0 I_n$, where $\alpha_0=10^{-3}$ and $I_n$ is the identity matrix with dimension $n$. 
\CORR{Because the covariance matrix converges within a few assimilation steps, the long-term estimates and the error metrics are insensitive to its initial value.}
The process noise covariance is set to $Q = \sigma_Q^2 I_n$ and the measurement noise covariance to $R = \sigma_R^2 I_m$, with the ratio $\beta=\sigma_Q^2/\sigma_R^2$ controlling the balance between model and measurement trust. The sensitivity of the filter performance to these parameters is assessed in Appendix~\ref{sec:sensitivity}. 
This isotropic, uncorrelated choice is relaxed in \S\ref{sec:estimated_cov}, where $Q$ and $R$ are instead estimated directly from DNS data.
At each assimilation step, the state transition $\boldsymbol{\Phi}$ is computed by integrating the ROM equations~\eqref{eq:ROM_compact} using a Runge-Kutta 5(4) scheme~\cite{dormand1980family} with adaptive step size. The state transition matrix $F_k$ is approximated via the first-order linearization in eq.~\eqref{eq:Jacobian}, where the analytical Jacobian of the ROM is used. The availability of a closed-form Jacobian, a natural consequence of the polynomial structure of Galerkin ROMs, makes the EKF implementation computationally efficient compared to finite-difference approximations.

\CORR{The ROM and the EKF have been implemented in Python using general-purpose ODE integrators, without dedicated performance optimization. As reported in the companion paper~\cite{flores2025galerkin}, even with this suboptimal numerical setup, model C96 achieves a 225-fold speed-up with respect to the DNS in the chaotic regime ($\R=Ra/Ra_c=120$, $Pr=10$). In the same conditions, one complete EKF cycle (prediction and update) with the C96 model and the coarse-grid strategy from \S\ref{sec:coarse_grid} ($m=48$) takes approximately 7.7\,ms. To place these numbers in a physical context, consider a water-filled cell at $Ra\approx 2\times10^5$, which corresponds to a layer depth of about $1$\,cm and a temperature difference of 10\,K. The convective time unit $h^2/(\kappa\sqrt{Ra})$ is then of order $1.56$\,s, so the assimilation interval used in this work, $\Delta t\simeq 0.19$, corresponds to roughly $296$\,ms of physical time. This is nearly forty times longer than the time required to process one assimilation step, meaning that, even with the present unoptimized implementation, real time operation of the current EKF-ROM framework is possible.
}

\section{Results}\label{sec:results}
The above-described state estimation framework is now applied to two-dimensional RB convection in different observation scenarios. If not stated otherwise, the ROM with $n=96$ modes (C96) is used in the KF prediction step. In \S\ref{sec:coarse_grid}, the filter performance is evaluated using a coarse grid of sensors across three dynamical regimes: periodic, quasiperiodic, and chaotic. It is then applied to \CORR{velocity-only} observations in \S\ref{sec:velocity_only}, a configuration representative of hybrid simulation methodologies. A greedy sensor placement algorithm is presented in \S\ref{sec:sensor_placement} to explore the optimization of sensor locations and to identify which measurement types contribute most to the estimation quality.
Finally, in \S\ref{sec:estimated_cov}  we examine the effect of replacing the isotropic noise covariances by data-estimated ones.

\subsection{Coarse grid sensing strategy}\label{sec:coarse_grid}
\begin{figure}[h]
	\centering
	\nolinenumbers
	\includegraphics[width=0.5\linewidth]{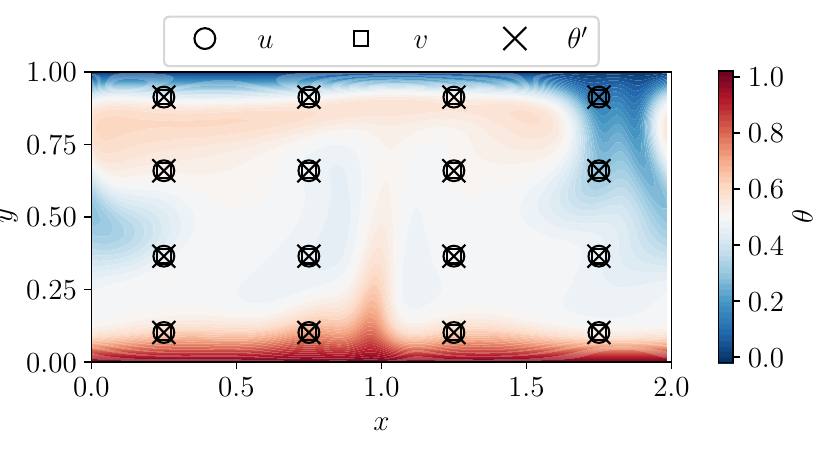}
	\caption{\CORR{Coarse-grid sensor layout. Symbols denote the variables measured at each probe: circles are used for the horizontal velocity, $u$, squares for the wall-normal component, $v$, and crosses for temperature, $\theta'$. The background displays an instantaneous DNS temperature field ($\mathcal{R} = 120$, $Pr = 10$).}}
	\label{fig:coarse_grid_sensing}
\end{figure}

Firstly, we apply the methodology to a simple sensing strategy consisting of a coarse grid of sensors.
Figure~\ref{fig:coarse_grid_sensing} illustrates the sensor arrangement in the $x,\,y$ plane, with the background showing an instantaneous temperature field from a DNS. 
A total of 16 probes arranged in a $4 \times 4$ grid are used. This grid is obtained by decimating the DNS mesh. 
Wall points are explicitly excluded since \CORR{direct velocity and temperature measurements at these locations} carry no information due to the no-slip and imposed-temperature boundary conditions. 
 Each probe measures the two velocity components and the temperature perturbation, $\theta^\prime$, yielding an observation space of dimension $m = 48$. In Fig.~\ref{fig:coarse_grid_sensing}, circles, squares, and crosses denote horizontal velocity, wall-normal velocity, and temperature sensors, respectively. The covariance matrices for the model and measurement noise are set to $Q = \beta I_n$ and $R = I_m$, respectively, where $\beta = 0.01$ is the model-to-measurement noise ratio. The sensitivity of the filter to $\beta$, the ROM dimension and the number of sensors is discussed in Appendix~\ref{sec:sensitivity}.

\begin{figure}[!h]
     \centering
     \nolinenumbers
    \includegraphics[width=1\linewidth]{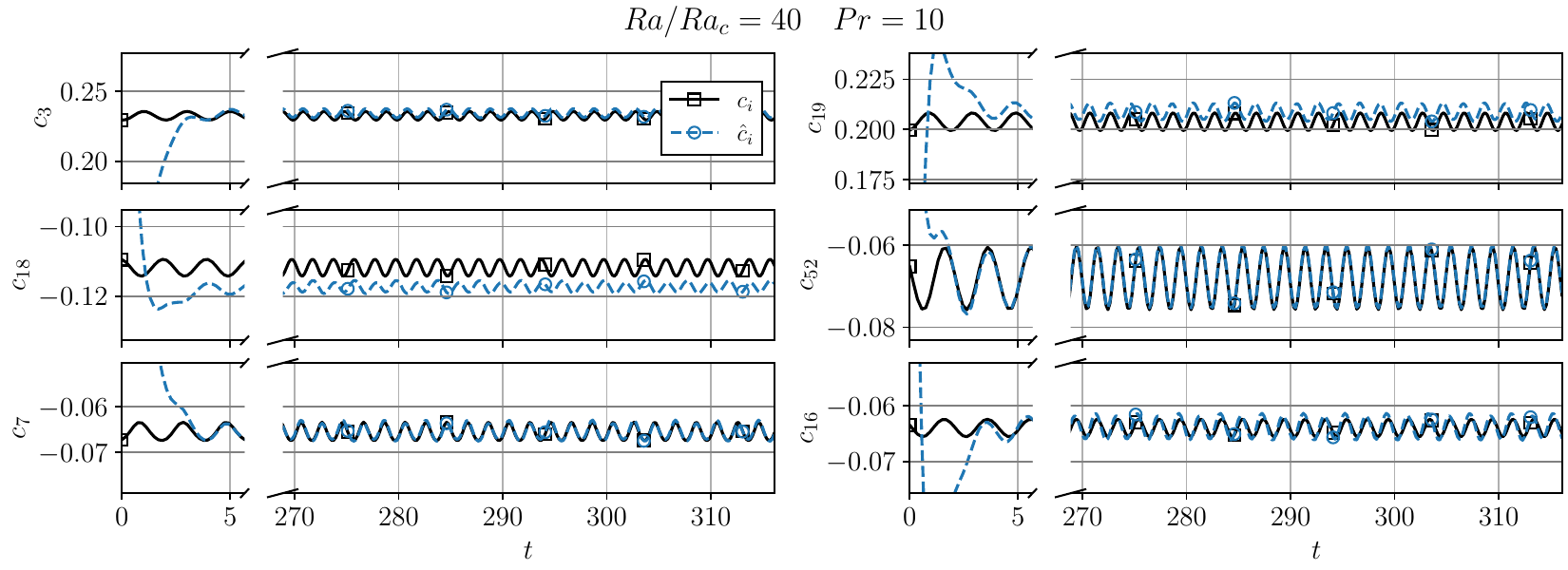}
    \caption{Time evolution of amplitude coefficients: comparison between DNS projection, $c_{i}$, and KF estimates, $\hat{c}_{i}$ for the six modes with the largest mean energy, $\langle c_i^2\rangle$.}
    \label{fig:coarse_grid_R040_Pr010_cik}
\end{figure}

The KF is run over the $\R=40$ and $Pr=10$ DNS. For this combination of \CORR{flow} parameters, the system is periodic, describing a stable orbit around the attractor. Figure~\ref{fig:coarse_grid_R040_Pr010_cik} shows the time evolution of the DNS coefficients, $c_{i,k}$ (black lines and squares), and their corresponding estimates, $\hat{c}_{i,k}$ (blue \CORR{dashed} lines and circles).  Only the six modes with the largest mean energy, $\langle c_i^2\rangle$,  are shown. \CORR{For this regime, these six modes account for $90.8\%$ of the total resolved energy.} The $x$-axis has been broken into two time segments to show the filter initialization and the last oscillation cycles of the simulation. During the filter initialization, the estimate converges towards the state within approximately five characteristic times. The overall agreement between the filter estimate and the state is very good with both signals nearly overlapping across the entire run. 
\CORR{For some coefficients, e.g. $c_{18}$, there is a steady offset between the filter estimate, $\hat{c}_i$, and the DNS projection, $c_i$. This suggests a systematic error in the truncated ROM whose attractor can be slightly shifted in state space with respect to the full-order system. While the Kalman update continuously corrects the trajectory using the incoming measurements, the prediction step keeps pulling the state back toward the shifted attractor of the ROM, resulting in a small mean offset.}

\begin{figure}[!h]
    \centering
    \nolinenumbers
    \includegraphics[width=1\linewidth]{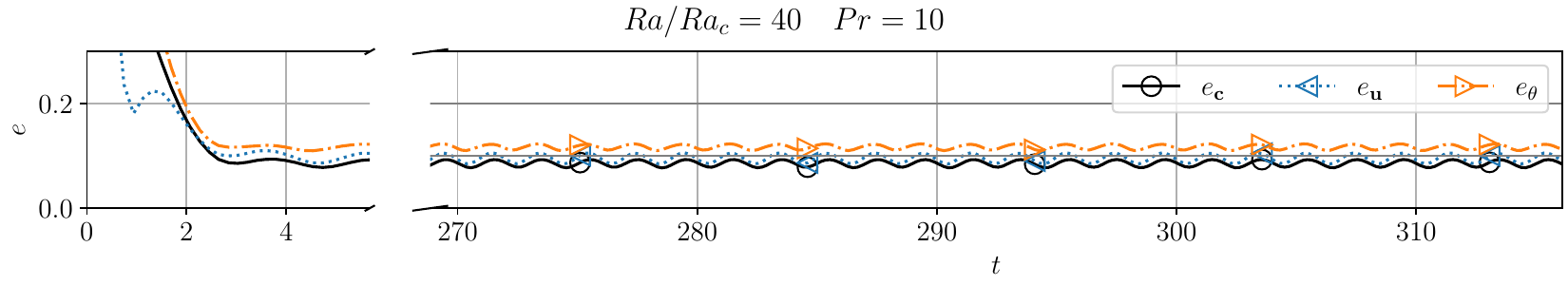}
    \caption{Time evolution of estimation, $e_\mathbf{c}$, velocity, $e_\mathbf{u}$, and temperature, $e_\theta$, errors.}
    \label{fig:coarse_grid_R040_Pr010_error}
\end{figure}

Figure~\ref{fig:coarse_grid_R040_Pr010_error} shows the evolution of the errors with time. After the initial transient, the filter error stabilizes around $8.9\%$, with small-amplitude oscillations around its mean value. 
\CORR{The velocity error oscillates around $E_\mathbf{u} \simeq 9.8\%$ and the temperature error stabilizes at  $E_\theta \simeq 12.0\%$.}

We now assess the influence of the dynamical regime on the filter performance by varying the Rayleigh number. In fluids with $Pr=10$, RB convection transitions to chaos via quasiperiodicity in the range $40<\R<120$. In our companion paper~\cite{flores2025galerkin}, it is shown that the current Galerkin ROMs are capable of predicting this bifurcation path. Here, the purpose is to evaluate whether the dynamical behavior has some impact on the performance of the Kalman filter. As the $Ra$ number is augmented, the scale separation grows, and the number of modes required to fully resolve the flow increases. Consequently, increasing $Ra$ renders the ROM a progressively coarser approximation of the full-order dynamics.

\begin{figure}[h]
	\centering
	\nolinenumbers
	\includegraphics[width=1\linewidth]{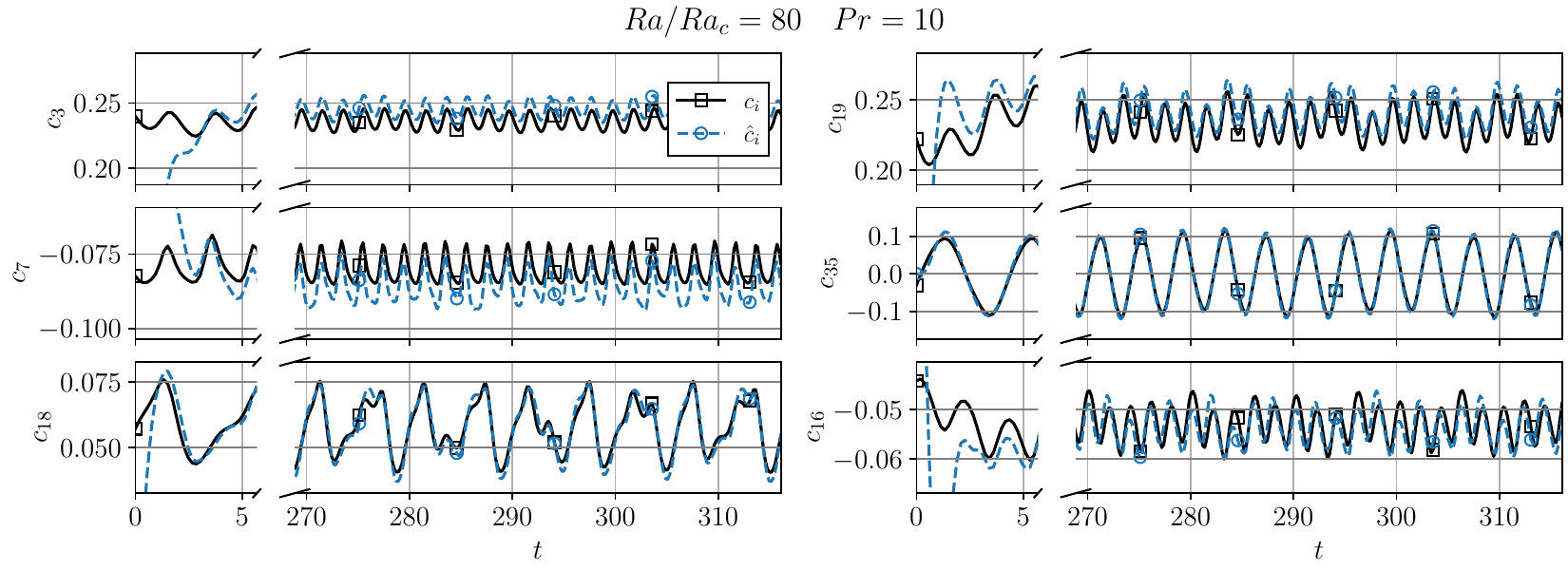}
	\caption{Time evolution of amplitude coefficients: comparison between DNS projection, $c_{i}$, and KF estimates, $\hat{c}_{i}$ for the six  modes with the largest mean energy, $\langle c_i^2\rangle$.}
	\label{fig:coarse_grid_coeffs_r080}
\end{figure}

 For $\R=80$ and $Pr=10$, the system is quasiperiodic. Figure~\ref{fig:coarse_grid_coeffs_r080} shows the evolution of the most energetic modes \CORR{(88.3\% of the total energy)} over time. Excellent agreement is also obtained for this $\R$ number with the filter estimates closely following the DNS signal. The Kalman filter successfully tracks the state in spite of the richer spectral content of the quasiperiodic attractor. For this $\R$ number, the instantaneous error signals (not shown here) display similar trends as in the periodic case though their mean values are slightly higher. 
 As for $\R=40$, the temperature error is the \CORR{highest in magnitude with a mean value of approximately $E_\theta=16.8\,\%$ followed by the velocity error, $E_\mathbf{u} = 12.9\%$, and the filter error, $E_\mathbf{c}=11.9\%$.} The moderate increase in all error metrics relative to the $\R=40$ case -- roughly $3\%$ in $E_\mathbf{c}$ and $E_\mathbf{u}$, and \CORR{nearly $5\%$ in $E_\theta$} -- is attributable to two concurrent effects: the increased dynamical complexity of the quasiperiodic attractor, and the larger truncation error of the ROM.

\begin{figure}[h]
    \centering
    \nolinenumbers
    \includegraphics[width=1\linewidth]{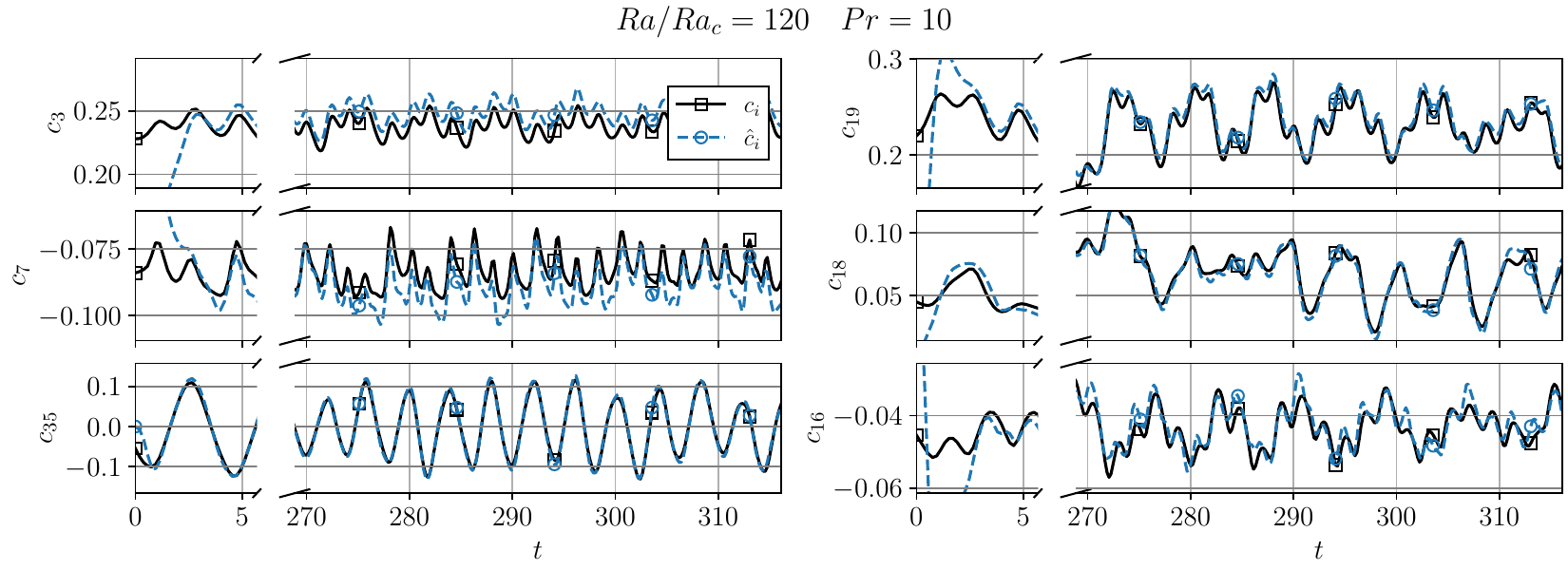}
    \caption{Time evolution of amplitude coefficients: comparison between DNS projection, $c_{i}$, and KF estimates, $\hat{c}_{i}$ for the six  modes with the largest  mean energy, $\langle c_i^2\rangle$.}
    \label{fig:coarse_grid_coeffs_r120}
\end{figure}

When the Rayleigh number is further increased to $\R=120$, the system becomes chaotic.  In this regime, the existence of positive Lyapunov exponents in the attractor introduces a fundamental challenge for state estimation: any initial uncertainty in the state grows exponentially in time. Therefore, the filter must continuously correct the diverging trajectory using incoming sensor data. Figure~\ref{fig:coarse_grid_coeffs_r120} shows the modal coefficient evolution for $\R=120$.  
\CORR{At this $\R$ number, the six most energetic modes carry $87.2\%$ of the total resolved energy.}
Despite the chaotic dynamics, there is a very good agreement between the DNS projection and the filter estimates. The filter successfully tracks the chaotic trajectory throughout the simulation window, with the estimates closely following the DNS signal. 

The instantaneous error signals at $\R=120$ are not shown here but are analogous to those of Fig.~\ref{fig:coarse_grid_R040_Pr010_error} at $\R=40$, except for their chaotic character and higher mean values. A summary of the filter errors as a function of the dynamical regime is presented in Tab.~\ref{tab:errors}. The error metrics for the chaotic case confirm a progressive degradation of filter performance with increasing $\R$. Nonetheless, the errors remain moderate even at $\R=120$, with $E_\mathbf{c} \approx 13.8\%$ and \CORR{$E_\theta \approx 18.7\%$}. This progressive increase is consistent with the combined effects of growing dynamical complexity and increasing ROM truncation at higher $\R$. Overall, these results demonstrate that the Kalman filter, combined with the Galerkin ROM, provides robust state estimation across all three dynamical regimes (periodic, quasiperiodic, and chaotic), despite relying on a fixed low-dimensional approximation and a sparse, suboptimal sensor configuration.

\begin{table}[h]
	\nolinenumbers
	\centering
	\begin{tabular}{ccccc}
		Rayleigh number & Dynamical regime & \begin{tabular}[c]{@{}c@{}}Filter error\\ $E_\mathbf{c}$ {[}\%{]}\end{tabular} & \begin{tabular}[c]{@{}c@{}}Velocity error\\ $E_\mathbf{u}$ {[}\%{]}\end{tabular} & \begin{tabular}[c]{@{}c@{}}Temperature error\\ $E_\theta$ {[}\%{]}\end{tabular} \\ \hline
		$\R=40$           & Periodic         & 8.88                                                                 & 9.75                                                                            & \CORR{12.03}                                                                           \\
		$\R=80$           & Quasiperiodic    & 11.87                                                                & 12.92                                                                           & \CORR{16.76}                                                                           \\
		$\R=120$          & Chaotic          & 13.76                                                                      &      16.02                                                                            &               \CORR{18.70}                                                                 
	\end{tabular}
	\caption{Mean filter, velocity and temperature errors as a function of the dynamical regime.}
	\label{tab:errors}
\end{table}
 
 \CORR{In the results above, the synthetic observations are extracted directly from the DNS fields so that measurements do not have noise contamination. The robustness of the estimator to measurement noise is assessed in Appendix~\ref{sec:noise}, where additive Gaussian noise with an amplitude of up to $20\%$ of the RMS value of each measured field ($\mathrm{SNR}=14$\,dB) is introduced in the observations of the chaotic case. Although the filter is not informed of the added noise, i.e.\ $Q$ and $R$ are kept unchanged, the errors increase only moderately: $E_\mathbf{c}$ grows from $13.8\%$ to $18.4\%$, $E_\mathbf{u}$ from $16.0\%$ to $20.7\%$ and $E_\theta$ from $18.7\%$ to $21.8\%$.}
 
 These results can be placed in the context of related works that combine POD-Galerkin ROMs with data assimilation methods for flow estimation, such as \citet{kikuchi2015assessment} for a cylinder wake and \citet{sommer2023estimating} for a centrifugal pump. Although direct quantitative comparisons are not possible due to the different flow configurations, physical models, and error definitions, a common feature of these prior studies is that they operate in periodic regimes and rely on ROMs that require either stabilization schemes or continuous EKF corrections to prevent divergence. In contrast, the present framework employs a controllability-mode ROM that remains stable without closure terms and is assessed across three dynamical regimes --- periodic, quasiperiodic, and chaotic --- demonstrating that the estimator remains effective as the dynamics become progressively more complex.
 
\subsection{State estimates from velocity measurements}\label{sec:velocity_only}

As discussed in \S\ref{sec:intro}, hybrid simulations combine experimental measurements with numerical solvers to reconstruct flow quantities that are difficult to measure directly~\cite{suzuki2012reduced,bauer2022assimilation,suzuki2009unsteady1,suzuki2009unsteady2}. The simulation acts as a physics-informed interpolator that fills spatial and temporal gaps in the experimental data while enforcing dynamical consistency~\cite{nisugi2004fundamental,hayase2015review,suzuki2012reduced}. A particularly attractive application in RB convection is the inference of the temperature field from velocity measurements. In experiments, velocity fields can be obtained non-intrusively from PIV measurements, which typically provide two-component velocity fields on a measurement plane. Conversely, temperature measurements require either intrusive probes such as thermocouples, which offer only pointwise information, or complex experimental techniques. The ability to reconstruct the temperature field from PIV data would reduce the need for temperature instrumentation. 

In this section, we use DNS results as synthetic observations and perform a dynamic state reconstruction from velocity measurements only, using the Kalman filter combined with the Galerkin ROM as estimator. The goal is to assess whether the filter can infer the system state, which includes the temperature field, from velocity data alone, and to explore the potential of this framework for hybrid simulation applications. 

\begin{figure}[h]
	\centering
	\nolinenumbers
	\includegraphics[width=0.5\linewidth]{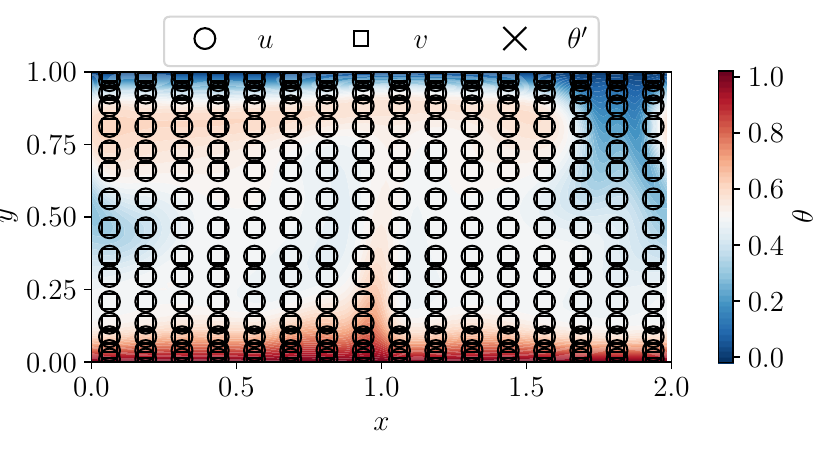}
	\caption{\CORR{Sensor configuration in the velocity-only observation strategy. Each probe measures the horizontal (circles) and the wall-normal (squares) velocity components. The background shows an instantaneous DNS temperature field at $\R = 120$, $Pr = 10$.}}
	\label{fig:UVgrid_sensing}
\end{figure}

\begin{figure}[h]
	\centering
	\nolinenumbers
	\includegraphics[width=1\linewidth]{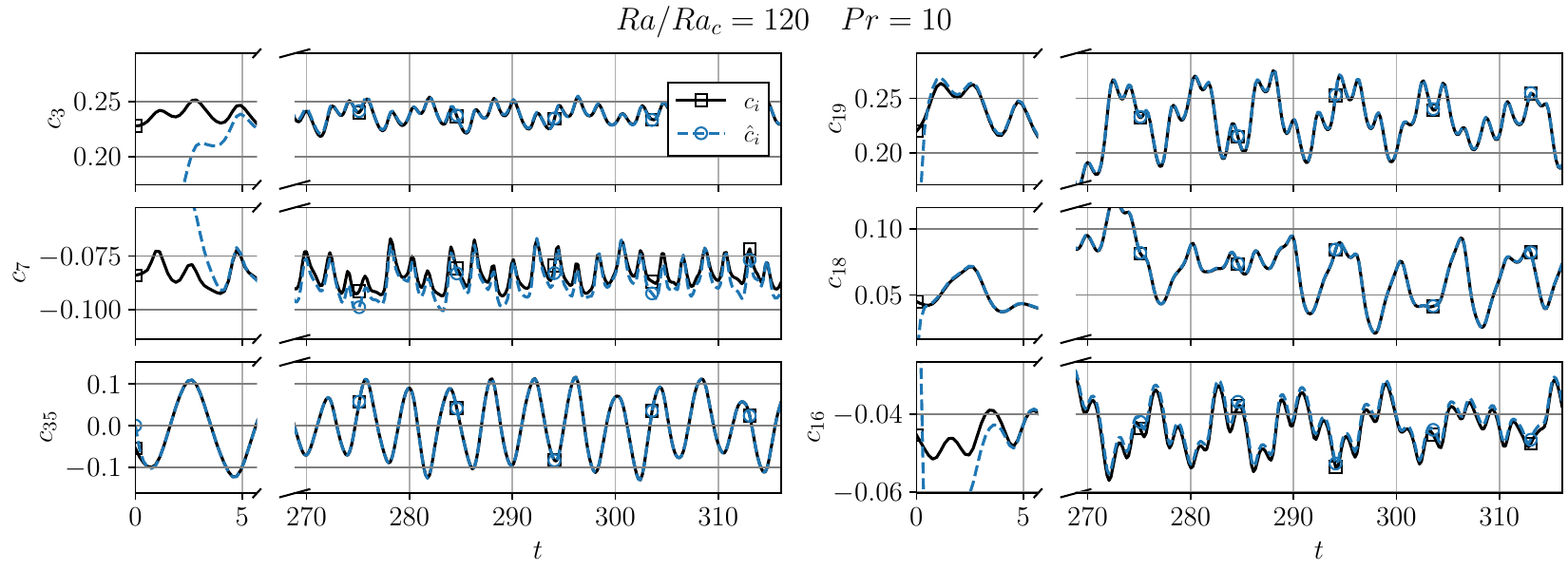}
	\caption{Time evolution of the dominant ROM coefficients for the velocity-only observation strategy at $\R = 120$, $Pr = 10$: DNS projection $c_i$ (black) and KF estimates $\hat{c}_i$ (blue).}
	\label{fig:UV_cik}
\end{figure}

To this end, the observation model is modified and reduced to velocity measurements only. A grid of $16 \times 16$ probes obtained by decimating the DNS mesh is distributed across the two-dimensional domain, yielding $n_u = n_v = 256$ measurement points. The sensor locations are shown in Fig.~\ref{fig:UVgrid_sensing} together with an instantaneous temperature field. At each probe location, the horizontal and vertical velocity components are measured, so that the observation vector becomes:
\begin{equation}
	\boldsymbol{y}_k = [u(\boldsymbol{x}_1, t_k),\,\ldots,\, u(\boldsymbol{x}_{n_u}, t_k),\, v(\boldsymbol{x}_1, t_k),\,\ldots,\, v(\boldsymbol{x}_{n_v}, t_k)]^T \in \mathbb{R}^{m},
\end{equation}
with $m = n_u + n_v = 512$ and $n_\theta = 0$. The observation matrix reduces to:
\begin{equation}
	H = \begin{bmatrix}
		H^u & H^v
	\end{bmatrix}^T \in \mathbb{R}^{m \times n},
\end{equation}
where $H^u$ and $H^v$ are the velocity sub-matrices defined in \S\ref{sec:Kalman}, and the temperature block $H^\theta$ is absent. All other filter parameters including the initial covariance and the process and measurement noise \CORR{covariances} are kept identical to the coarse-grid configuration described in \S\ref{sec:coarse_grid}.
The filter is evaluated for the chaotic regime at $\R = 120$ and $Pr = 10$ using the DNS with \CORR{random seed} $\eta=1$. Figure~\ref{fig:UV_cik} shows the time evolution of the dominant modal amplitudes. The DNS projection coefficients and the filter estimates are in excellent agreement, with only mode $c_7$ exhibiting minor discrepancies. These results indicate that velocity-only observations on a decimated grid are sufficient to track the system trajectory, even in the chaotic regime.

\begin{figure}[h]
	\centering
	\nolinenumbers
	\includegraphics[width=1\linewidth]{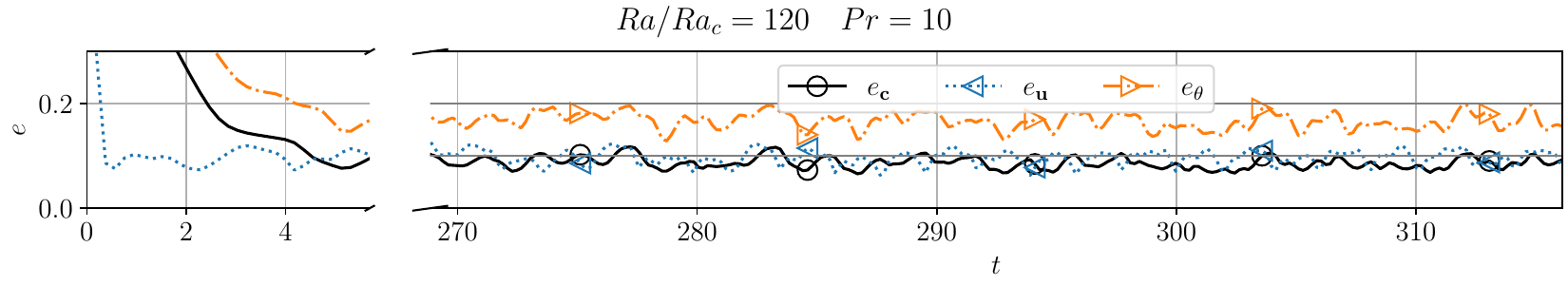}
	\caption{Time evolution of filter, velocity and temperature errors for the velocity-only observation strategy at $\R = 120$, $Pr = 10$.}
	\label{fig:UV_error}
\end{figure}

\CORR{The corresponding error signals are shown in Fig.~\ref{fig:UV_error}. The mean filter and velocity errors are $9.2$\% and $10.0$\%, respectively while the mean temperature error remains at roughly $17.3\%$.} This is attributed to the absence of a direct correction based on local temperature differences which results in a higher relative uncertainty in the reconstructed temperature. 

It is interesting to place these error levels in the context of related works. \citet{bauer2022assimilation} applied nudging to a 3D DNS using tomographic PIV data at $Ra = 7 \times 10^9$ and $Pr = 18$, with measured velocity fields at a resolution of $119 \times 119 \times 57$. The spatial correlations between mean reconstructed and measured fields reached $C_{T\tilde{T}} = 0.84$ for temperature and exceeded $0.9$ for velocity.
\citet{cornick2009state} applied a local ensemble transform Kalman filter (LETKF) to chaotic RB convection in a cylindrical domain at $Pr \approx 1$ and $\R=2$. They used dense shadowgraph observations ($\sim 1.5 \times 10^5$ pixels) sampled at $3$~Hz. In numerical experiment tests, their method achieved minimum temperature errors of $1$--$2\%$ and mean-flow errors of $2$--$3\%$, propagating $18$ ensemble members on a grid of $\sim 6 \times 10^5$ unknowns.
\citet{agasthya2022reconstructing} applied nudging to 2D DNS at $Ra = 7.2 \times 10^7$--$3.6 \times 10^8$ and $Pr = 1$, using temperature sensor grids ranging from $36$ to $14\,792$ measurement points. Temperature synchronization was achieved when the probe spacing was reduced to a few Kolmogorov lengths, but the velocity reconstruction saturated at a global $L^2$-error of approximately $40\%$ at the higher $Ra$, even when the full temperature field was provided.
\CORR{Establishing a direct quantitative comparison with these works is difficult because they differ from ours in dimensionality, Rayleigh number, type of data, observed variables and error metric. The comparison suggests that there is a trade-off between computational cost and accuracy. Full-order methods can in principle recover all scales of the flow, but require propagating one or several full-order simulations online and, in the case of nudging, observations resolving the small scales to synchronise with the true flow. The ROM-based EKF uses a much cheaper model and its predictions can be corrected with observations substantially coarser than the DNS grid. However, its accuracy is limited by the ROM truncation error and it requires a model valid in the regime of interest.}

\CORR{Overall, the present results demonstrate that, in a 2D configuration at moderate $Ra\approx 2\times 10^5$, low-resolution velocity measurements -- with a spatial resolution significantly coarser than the DNS grid -- are sufficient to reconstruct the instantaneous velocity field within $10\%$ error and the temperature field with a 17.3\% error.} 
The successful inference of the thermal state from velocity observations alone is enabled by the Galerkin ROM, which provides a sufficiently accurate representation of the coupled dynamics to compensate for the absence of direct temperature measurements.
\CORR{It must be noted that the observation strategy employed here does not fully replicate the characteristics of PIV measurements. In a real PIV setup, velocity vectors are typically mapped to a uniform grid and represent spatially window-averaged values. Here, the grid is non-uniform and measurements are pointwise. Incorporating more realistic sensor models with spatial filtering is a key step towards the experimental deployment of this technique. Furthermore, extending this framework to three-dimensional configurations, higher Rayleigh numbers, and noisy data remain promising directions for future work.}

\subsection{Greedy probe removal for sensor placement}\label{sec:sensor_placement}

In \S\ref{sec:coarse_grid}, it has been shown that  a coarse grid of measurement probes can provide accurate state estimates across different dynamical regimes. However, a natural question is whether all probes contribute equally to the estimation quality, or whether some can be removed without significant loss of accuracy. In this section, we present a sensor \CORR{placement strategy} that exploits the Kalman gain matrix to progressively identify and remove the least informative measurement locations. 

The Kalman gain matrix $K_k \in \mathbb{R}^{n \times m}$ computed during the update step provides a measure of the relevance of each measurement channel: it quantifies how much each observation influences the state correction at each time step. A large gain associated with a particular measurement indicates that the filter relies heavily on that observation to correct its prediction, while a gain near zero signals that the measurement is redundant given the model prediction and the other observations. This motivates the use of the Kalman gain to rank sensors and purge the least relevant.

To assess the long-term relevance of each measurement channel, the time-averaged norm of the Kalman gain is computed. Given the gain matrices $K_k$ for $k = 1, \ldots, N_\mathrm{it}$, the relevance score for the $j$-th measurement channel is defined as:
\begin{equation}\label{eq:gain_norm}
	\mathcal{S}_j = \sqrt{\sum_{k=1}^{N_\mathrm{it}} \sum_{i=1}^{n} K_{k,ij}^2},
\end{equation}
where $K_{k,ij}$ denotes the $(i,j)$-th entry of the Kalman gain at time step $k$. This score aggregates the contribution of the $j$-th observation across all state components and all assimilation steps, providing a ranking of sensor importance.

The sensor \emph{optimization} proceeds iteratively. The algorithm starts from an initial distribution of measurements with $n_u$ horizontal velocity, $n_v$ vertical velocity, and $n_\theta$ temperature sensors, yielding an initial output dimension $m = n_u + n_v + n_\theta$. At each iteration, the following steps are performed:
\begin{enumerate}
	\item \textbf{Run the EKF.} The EKF is run for $N_\mathrm{it}=1666$ assimilation steps using the current set of observations and the total filter error, $E_\mathbf{c}$, is computed. At each time step, the Kalman gain $K_k$ is recorded.
	\item \textbf{Remove the least relevant sensor}. Compute measurement relevance scores using eq.~\eqref{eq:gain_norm} and remove the least relevant sensor from the active set, reducing the output dimension by one. 
	\item \textbf{Update the observation matrix.} The observation matrix $H$ and the measurement noise covariance $R$ are rebuilt to reflect the reduced probe configuration.
\end{enumerate}

This greedy procedure is repeated until either the total error or the output dimension reach a prescribed value ($m=12$). At each iteration, the sensor configuration and the corresponding error, $E_\mathbf{c}$, are stored. The algorithm produces a sequence of progressively sparser observation strategies, each associated with an estimation error. By ranking probes across all three physical variables, the algorithm balances the number of velocity and temperature measurements, allowing the filter to retain the most informative combination regardless of the variable type. It is worth noting that this algorithm does not guarantee a globally optimal sensor configuration. Nevertheless, it provides a physically interpretable approach that leverages the information already produced by the KF at no additional cost. This iterative procedure is made tractable by the low computational cost of the ROM-based EKF, which allows multiple successive filter runs in reasonable times. 

The greedy removal algorithm has been applied to the chaotic case ($\R=120$ and $Pr=10$) using the simulation with random seed $\eta=1$. The algorithm is initialized from a homogeneous $8 \times 4$ sensor grid obtained from the decimation of the DNS grid. As in \S\ref{sec:coarse_grid}, each probe initially measures the two velocity components and the temperature perturbation, so that the initial output dimension is $m=96$. During the removal process, the KF is evaluated using model C96 and $\beta=0.01$. The initialization of covariance matrix and filter estimates remains unchanged with respect to \S\ref{sec:coarse_grid}.

\begin{figure}[h]
	\centering
	\nolinenumbers
	\begin{minipage}[c]{0.64\linewidth}
		\centering
		\includegraphics[width=\linewidth]{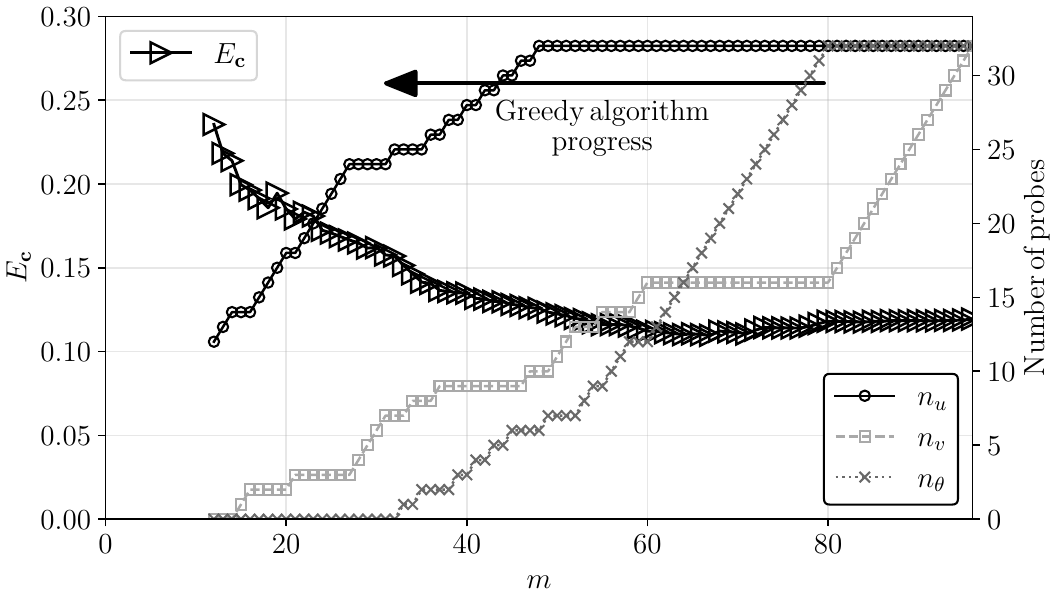}
		\subcaption{Total filter error $E_\mathbf{c}$ (left axis) and number of probes per variable (right axis) as a function of the output dimension $m$. Solid, dashed, and dotted lines indicate the number of horizontal velocity, $u$, wall-normal velocity, $v$, and temperature, $\theta$, probes, respectively. \CORR{The arrow indicates the greedy algorithm progress direction.}}
		\label{fig:greedy_error}
	\end{minipage}
	\hfill
	\begin{minipage}[c]{0.35\linewidth}
		\centering
		\includegraphics[width=\linewidth]{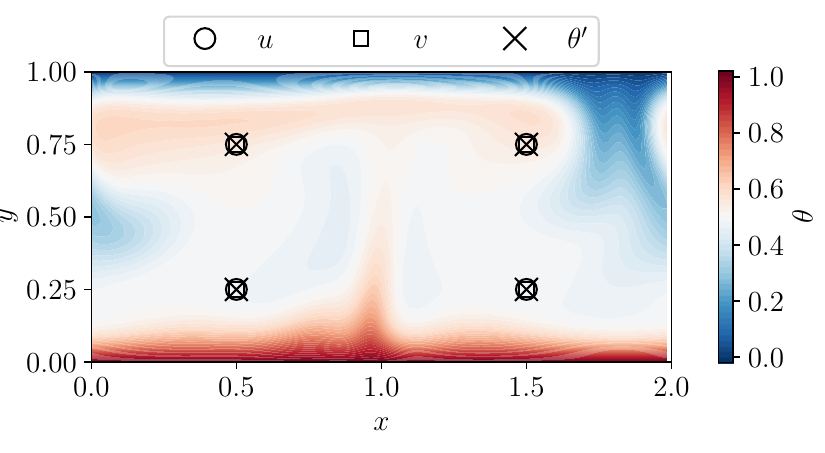}
		\subcaption{Coarse grid sensor layout ($m = 12$).}
		\label{fig:coarse_grid_sensing_m12}
		\vspace{4pt}
		\includegraphics[width=\linewidth]{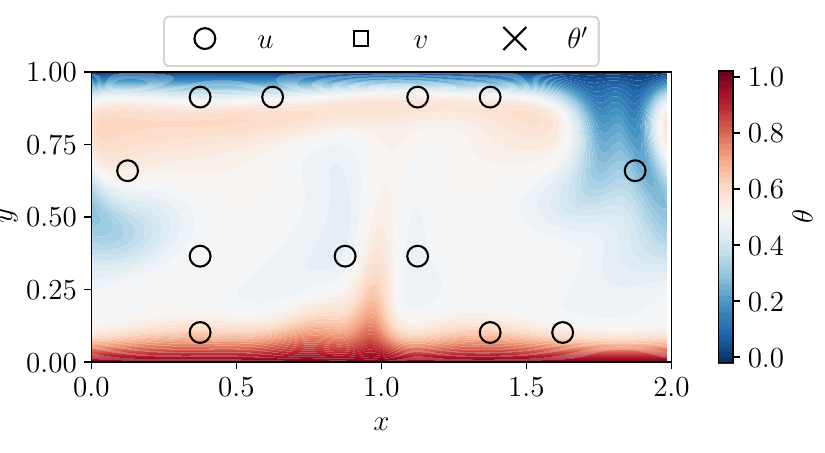}
		\subcaption{Greedy sensor layout ($m = 12$).}
		\label{fig:greedy_sensing_m12}
	\end{minipage}
	\caption{Greedy sensor removal results at $\R = 120$, $Pr = 10$, $\eta=1$. (a) Evolution of filter error and sensor count during the removal process. \CORR{The greedy algorithm advances from right to left (from $m=96$ to $m=12$), removing one sensor at each step}. (b,\,c) Resulting probe configurations for the coarse-grid and greedy strategies at $m = 12$. Circles, squares, and crosses denote horizontal velocity, wall-normal velocity, and temperature sensors, respectively. The background shows an instantaneous DNS temperature field.}
	\label{fig:greedy_overview}
\end{figure}

In Fig.~\ref{fig:greedy_error}, the evolution of the total filter error, $E_\mathbf{c}$, is shown on the left axis as a function of the observation dimension, $m$, while the right axis displays the number of sensors by type. 
\CORR{The greedy algorithm progresses from right to left, starting from the initial configuration of $m=96$ sensors and ending at $m=12$, by removing one sensor at each step.} As sensors are progressively removed, the total filter error first decreases slightly suggesting that some measurements are not only redundant but mildly detrimental to the estimate. Then, the error increases monotonically, reaching roughly $24$\% for $m = 12$. This increase remains moderate at first and accelerates as the observation dimension shrinks further. Interestingly, the algorithm preferentially removes temperature probes first, followed by vertical velocity sensors, retaining horizontal velocity measurements as the most informative channels.  This hierarchy indicates that, for the configuration considered, horizontal velocity observations carry the greatest corrective weight in the Kalman update. The specific arrangement of the retained probes varies if the greedy algorithm is run using different filter parameters or DNS realizations. 
However, the overall ranking of variable types remains consistent across runs, with horizontal velocity measurements being systematically retained the longest.

We now compare the sensor configuration resulting from the greedy algorithm with $m = 12$ against a coarse-grid strategy \CORR{with} the same observation dimension. As in previous examples, the coarse grid is obtained by decimating the DNS grid into a $2 \times 2$ arrangement, measuring both velocity components and the temperature at each grid point. Fig.~\ref{fig:coarse_grid_sensing_m12} shows the resulting probe layout for this coarse-grid strategy, while Fig.~\ref{fig:greedy_sensing_m12} displays the configuration produced by the greedy algorithm. Note that in the latter, only horizontal velocity sensors have been retained. The spatial sensor distribution in the greedy algorithm does not seem to follow a clear geometric pattern, suggesting that the algorithm exploits flow-specific features.

\begin{figure}[h]
	\centering
	\nolinenumbers
	\includegraphics[width=1\linewidth]{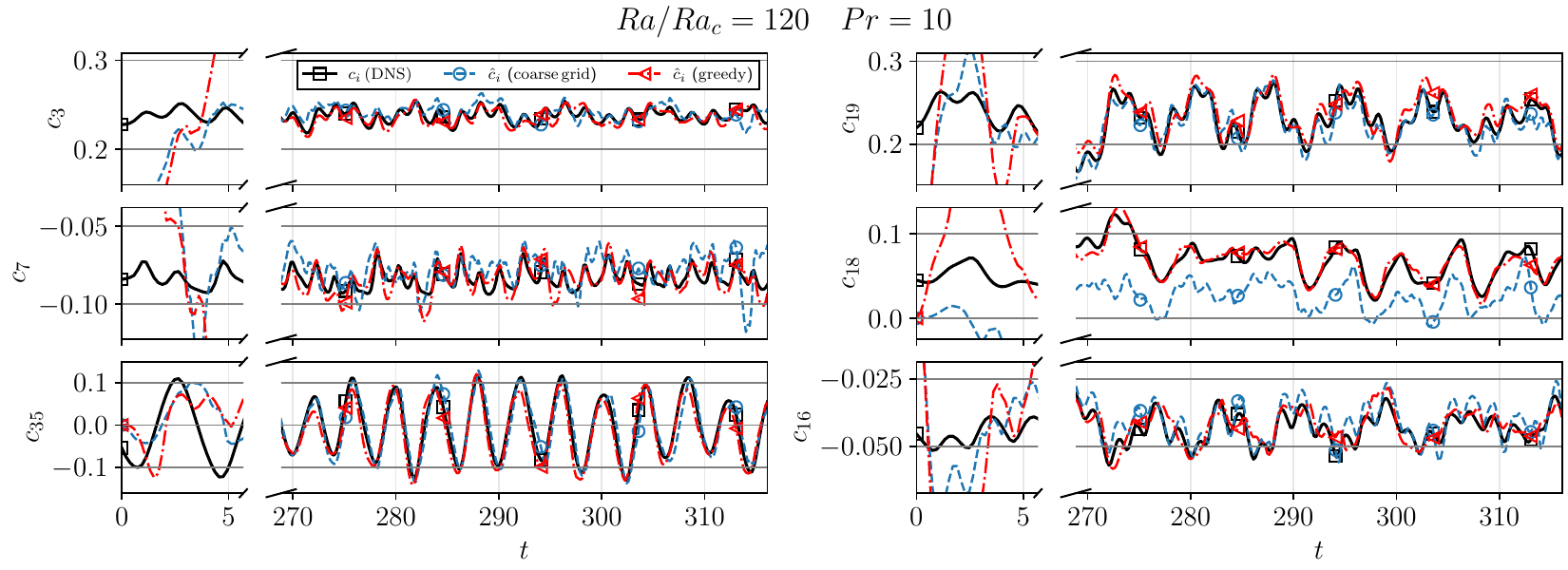}
	\caption{\CORR{Time evolution of the most energetic modes for the chaotic case ($\R = 120$, $Pr = 10$, $\eta = 1$): DNS projection (black solid lines), coarse-grid strategy estimates with $m=12$ (blue dashed lines), and greedy strategy estimates with $m=12$ (red dash-dot lines).}}	
	\label{fig:greedy_vs_coarse_12_coeffs_r120}
\end{figure}

\CORR{The comparison between the filter estimates and the DNS projection for the most energetic modes is shown in Fig.~\ref{fig:greedy_vs_coarse_12_coeffs_r120}. The DNS projection is represented by black solid lines, the coarse-grid estimates by blue dashed lines, and the greedy strategy estimates by red dash-dot lines.}
Both strategies yield reasonable tracking of the dominant modes, with the filter estimates following the DNS projection. The most notable difference appears for mode $c_{18}$, where the greedy strategy clearly outperforms the regular grid. The improvement in performance obtained with the greedy algorithm is better reflected in the global error metrics: the coarse-grid yields $E_\mathbf{u} = 84.7\%$, \CORR{$E_\theta = 31.3\%$}, and $E_\mathbf{c} = 64.0\%$, whereas the greedy arrangement reduces these to $E_\mathbf{u} = 26.9\%$, \CORR{$E_\theta = 25.7\%$}, and $E_\mathbf{c} = 24.0\%$. The greedy strategy thus achieves a substantially lower total error despite using the same observation dimension, with the improvement being particularly pronounced for the velocity field.

\begin{figure}[h]
	\centering
	\nolinenumbers
	\includegraphics[width=0.6\linewidth]{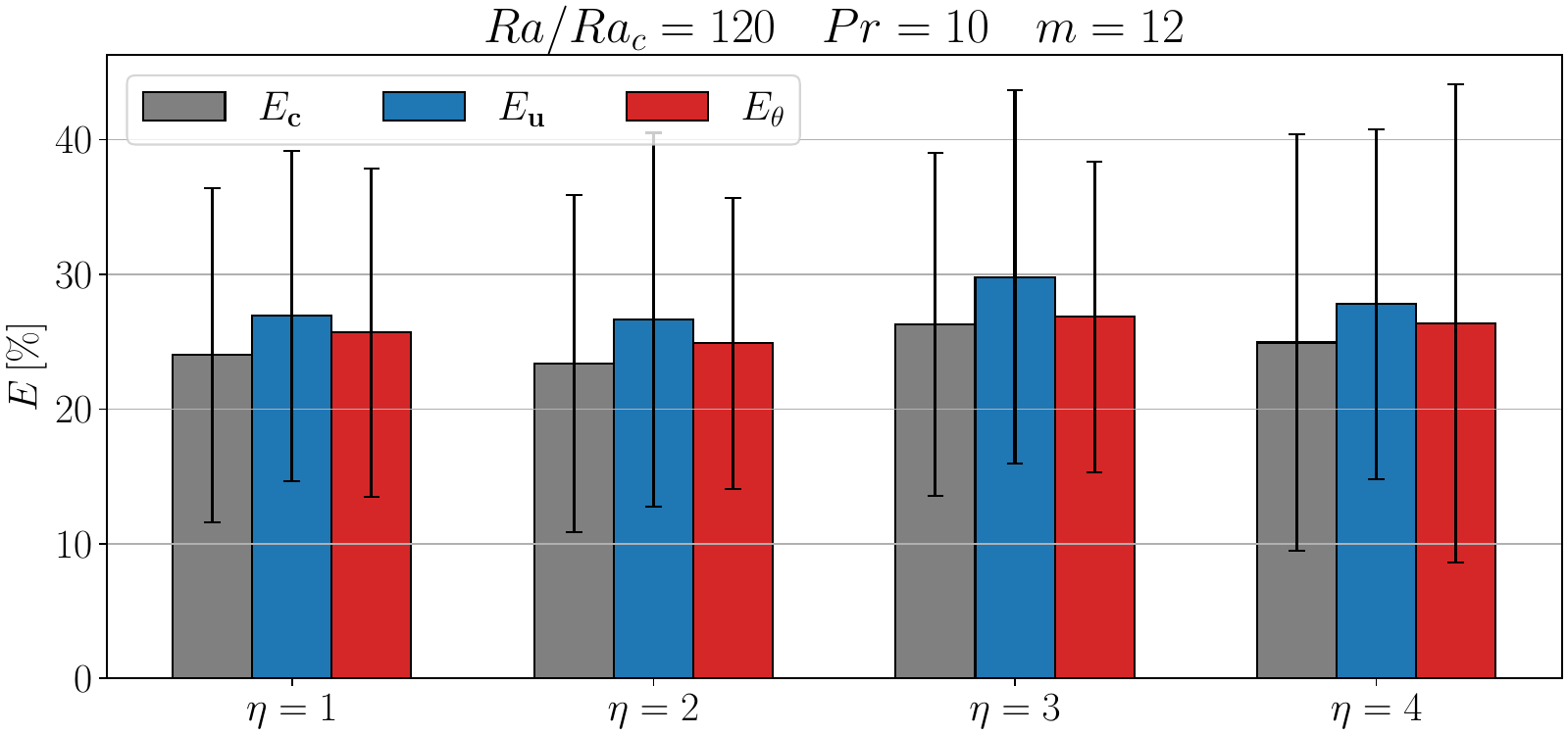}
	\caption{Error metrics $E_\mathbf{u}$, $E_\theta$, and $E_\mathbf{c}$ for the greedy strategy ($m = 12$) across four DNS realizations with different random seeds $\eta$. Error bars represent the \CORR{standard deviation of the instantaneous error signals.}}
	\label{fig:rseed_comparison}
\end{figure}

A natural question is whether the greedy observation strategy generalizes beyond the specific DNS realisation used to derive it, or whether its performance gains are tied to the particular flow features of that run. To address this, the Kalman filter has been applied using the greedy observation strategy with $m=12$ to three additional DNS realizations with different random initial conditions. The results are presented in Fig.~\ref{fig:rseed_comparison}, where the three error metrics are displayed for each random seed $\eta$. The greedy strategy was derived using $\eta = 1$; the figure shows that comparable error levels are obtained across all four realizations, despite the coherent structures, temperature plumes, and flow vortices occupying different spatial locations in each simulation. 

\begin{figure}[h]
	\centering
	\nolinenumbers
	\begin{subfigure}[b]{0.49\linewidth}
		\centering
		\includegraphics[width=\linewidth]{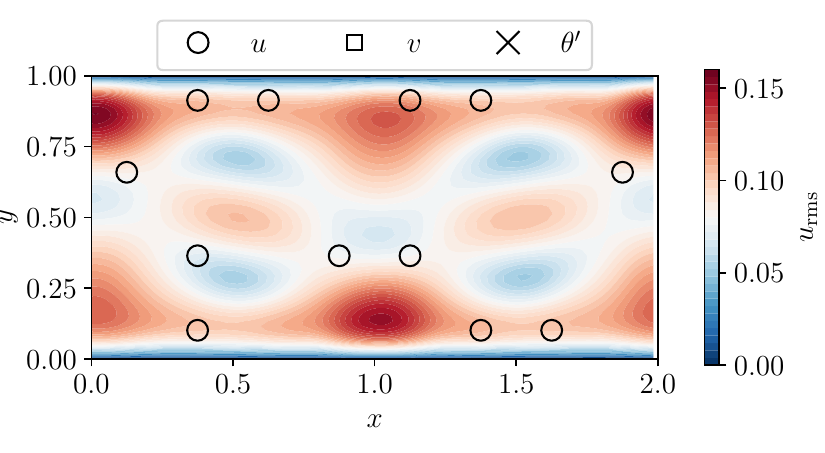}
		\caption{$\R=120$, $Pr=10$, $\eta=1$.}
		\label{fig:urms_rseed01}
	\end{subfigure}
	\begin{subfigure}[b]{0.49\linewidth}
		\centering
		\includegraphics[width=\linewidth]{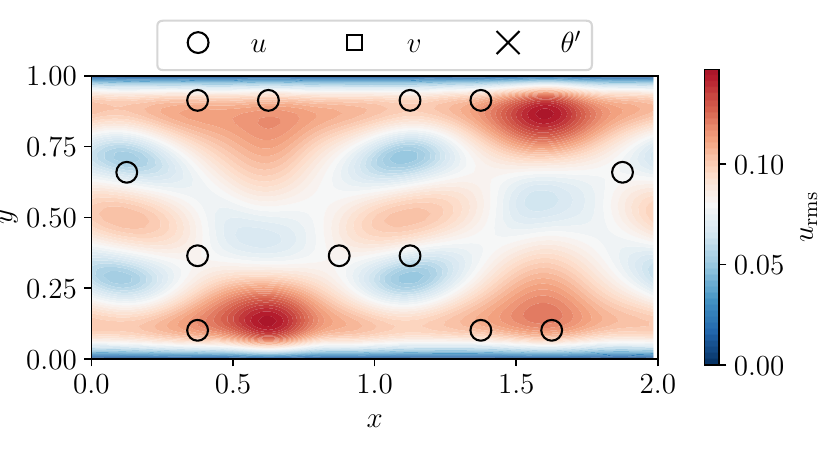}
		\caption{$\R=120$, $Pr=10$, $\eta=2$.}
		\label{fig:urms_rseed02}
	\end{subfigure}
	\begin{subfigure}[b]{0.49\linewidth}
		\centering
		\includegraphics[width=\linewidth]{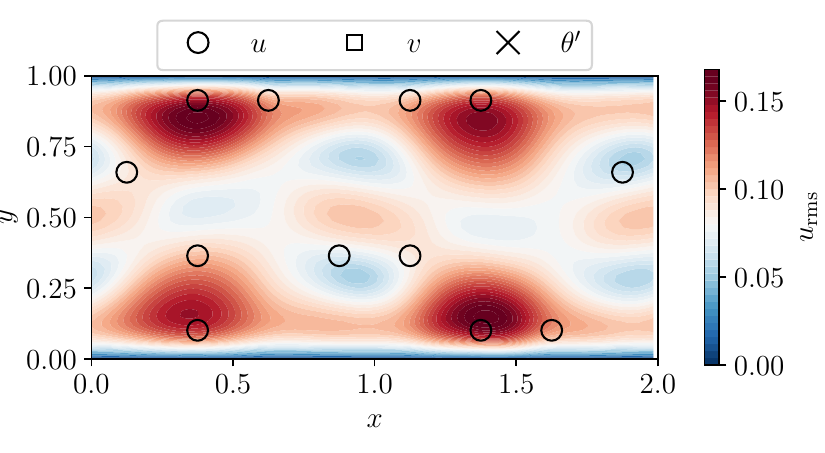}
		\caption{$\R=120$, $Pr=10$, $\eta=3$.}
		\label{fig:urms_rseed03}
	\end{subfigure}
	\begin{subfigure}[b]{0.49\linewidth}
		\centering
		\includegraphics[width=\linewidth]{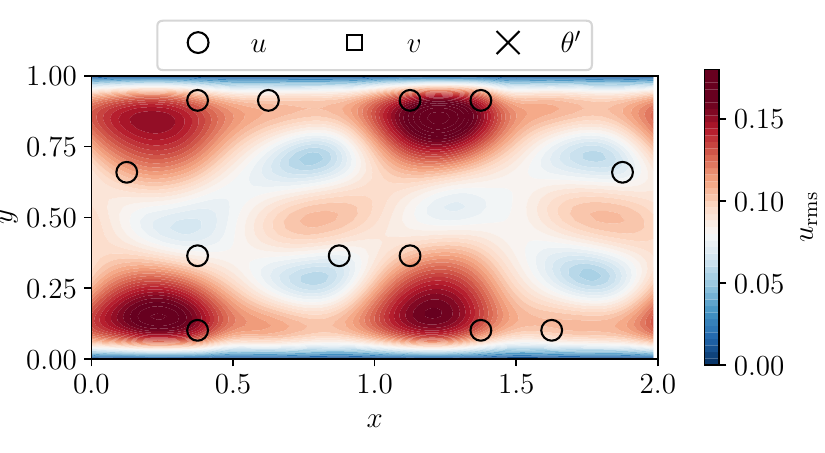}
		\caption{$\R=120$, $Pr=10$, $\eta=4$.}
		\label{fig:urms_rseed04}
	\end{subfigure}
	\caption{Greedy sensor configuration at $m = 12$ overlaid on the root-mean-square horizontal velocity $u_\mathrm{rms}$ for four DNS realizations with different initial conditions. Symbols indicate the retained probe locations.}
	\label{fig:urms_rseeds}
\end{figure}

To shed light on the preferential locations for sensor placement, Fig.~\ref{fig:urms_rseeds} overlays the greedy sensor configuration at $m = 12$ with the root-mean-square horizontal velocity field $u_\mathrm{rms}$ computed from four DNS realizations with different initial random seeds. The $u_\mathrm{rms}$ fields are characterized by moderate velocity fluctuation values close to the domain walls, right after the boundary layers. A significant part of the retained probes tend to cluster in these regions of high velocity fluctuations, suggesting that the greedy algorithm identifies dynamically active zones as the most informative measurement locations. Nonetheless, five of the twelve sensors are retained in the central region of the domain, indicating that measurements of the horizontal velocity away from the walls are also necessary to adequately correct the model estimates.

The greedy sensor placement strategy presented here achieves substantially lower estimation errors than a coarse-grid arrangement of the same observation dimension. It also reveals that horizontal velocity measurements carry the greatest corrective weight in the Kalman update. The retained sensors tend to cluster in regions of moderate-to-high velocity fluctuations near the walls, indicating that dynamically active zones are the most informative measurement locations. The low computational cost of the ROM-based EKF makes the iterative removal procedure tractable and enables a significant reduction of the observation dimension without excessive loss of accuracy. 
\CORR{Because greedy algorithms are path-dependent, the final sensor layout depends on the initial grid density and the algorithmic parameters such as the removal rate. However, it provides a physically interpretable baseline and the global trends remain consistent: preference for horizontal velocity measurements and sensor clustering in dynamically active regions.} Future work could explore more formal optimization techniques, such as convex relaxation or combinatorial search methods, to further refine sensor placement.

\subsection{Data-estimated noise covariances}\label{sec:estimated_cov}

Throughout the preceding sections, the process and measurement noise covariances were taken as isotropic and uncorrelated, $Q = \beta I_n$ and $R = I_m$. This is a common and convenient choice in many practical implementations of \CORR{the} KF. However, this choice embeds two assumptions: that every mode/measurement channel has the same error (isotropy), and that the errors in different modes/channels are uncorrelated (a diagonal noise matrix). Neither of these assumptions necessarily holds in practice. 
\CORR{To assess the impact of this simplification, DNS results can be used to estimate the model and measurement noise covariances. Comparing these estimates against the baseline model reveals the structure of the matrices and permits quantifying the limitations of the isotropic approach.}

The model-error covariance, $\hat{Q}$, can be estimated from the one-step prediction discrepancy of the ROM. For each snapshot, the ROM is advanced a single step from the DNS-projected state $\boldsymbol{c}_k$, yielding $\tilde{\boldsymbol{c}}_{k+1} = \boldsymbol{\Phi}(\boldsymbol{c}_k; Ra, Pr, \Delta t_k)$, and the residual $\boldsymbol{\varepsilon}_k = \boldsymbol{c}_{k+1} - \tilde{\boldsymbol{c}}_{k+1}$ is recorded. Seeding the integration from the true (DNS-projected) state allows isolating the model error. The sample covariance over the $N_s$ residuals in the assimilation window gives the model-error covariance:
\begin{equation}\label{eq:Qhat}
	\hat{Q} = \frac{1}{N_s-1}\sum_{k=1}^{N_s} (\boldsymbol{\varepsilon}_k - \bar{\boldsymbol{\varepsilon}})(\boldsymbol{\varepsilon}_k - \bar{\boldsymbol{\varepsilon}})^{T}
	\qquad \text{with}\quad\bar{\boldsymbol{\varepsilon}} = \frac{1}{N_s}\sum_{k=1}^{N_s}\boldsymbol{\varepsilon}_k.
\end{equation}
Analogously, the measurement-error covariance can be estimated from the off-projection residual at the probes, $\boldsymbol{r}_k = \boldsymbol{y}_k - H\boldsymbol{c}_k$,
\begin{equation}\label{eq:Rhat}
	\hat{R} = \frac{1}{N_s-1}\sum_{k=1}^{N_s} (\boldsymbol{r}_k - \bar{\boldsymbol{r}})(\boldsymbol{r}_k - \bar{\boldsymbol{r}})^{T}\qquad \text{with}\quad\bar{\boldsymbol{r}} = \frac{1}{N_s}\sum_{k=1}^{N_s}\boldsymbol{r}_k.
\end{equation}

We have computed the noise covariances from the DNS with $\R=120$, $Pr=10$ and \CORR{random seed} $\eta=1$, using the observation matrix, $H$, of the coarse-grid strategy with $m=48$ sensors described in \S\ref{sec:coarse_grid}. The filter is then run for the three dynamical regimes with the estimated matrices, $Q = \hat{Q}$ and $R = \hat{R}$, in place of the isotropic ones. We deliberately reuse the chaotic-case covariances to evaluate the periodic and quasiperiodic regimes as well. This is done to test whether a single estimate transfers across dynamical regimes. The structure of the estimated noise matrices is analyzed in detail in Appendix~\ref{sec:cov_structure}.

\begin{figure}[h]
	\centering
	\nolinenumbers
	\includegraphics[width=1\linewidth]{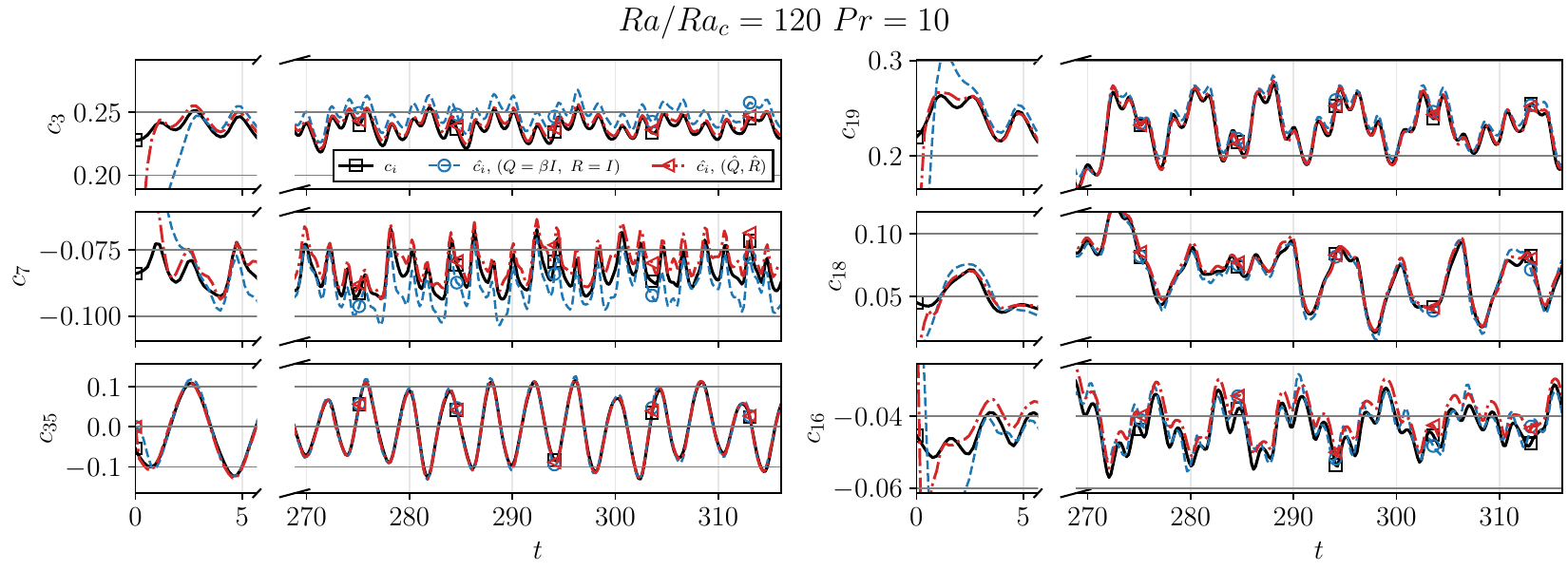}
	\caption{Time evolution of the most energetic modes for the chaotic case ($\R = 120$, $Pr = 10$, $\eta = 1$, $m = 48$): DNS projection $c_i$ (black solid lines), baseline isotropic filter with $Q = \beta I$, $R = I$ (blue dashed lines), and filter with data-estimated covariances, $\hat{Q}$, $\hat{R}$ (red dash-dot lines).}
	\label{fig:estcov_comparison}
\end{figure}

Figure~\ref{fig:estcov_comparison} shows the modal-amplitude tracking for the most energetic modes in the chaotic case ($\R = 120$, $Pr = 10$, $\eta = 1$) with the $m = 48$ coarse-grid observation strategy. Blue dashed lines denote the filter estimates from the baseline isotropic covariances, and red dash-dot lines those from the data-estimated noise matrices. Both filters track the dominant modes, but the EKF with the estimated covariance matrices yields a closer fit. The improvement in performance is better reflected in the error metrics: the filter error decreases from 13.8\% to 9.2\%, the velocity error from 16.0\% to 12.9\% and the temperature error from \CORR{18.7\% to 15.3\%}. 
Table~\ref{tab:cov_comparison} reports the global error metrics across all three dynamical regimes. 
Using the estimated covariance matrices improves every error metric in every regime with the largest benefits in the chaotic regime, the case that was used to estimate the noise matrices. 
\CORR{Calibrating these matrices requires access to the true state of the system, $\boldsymbol{c}_k$, which is generally unavailable in experiments. However, an offline DNS could be used to perform this calibration \emph{a priori} to enhance the filter performance. }

\begin{table}[h]
	\centering
	\nolinenumbers
	\begin{tabular}{cc|c|c|c|c|c|c}
		& & \multicolumn{2}{c|}{$E_\mathbf{c}$ [\%]} & \multicolumn{2}{c|}{$E_\mathbf{u}$ [\%]} & \multicolumn{2}{c}{$E_\theta$ [\%]} \\
		$\R$ & Regime & baseline & $\,\,\,\hat{Q},\hat{R}\,\,\,$ & baseline & $\,\,\,\hat{Q},\hat{R}\,\,\,$ & baseline & $\,\,\,\hat{Q},\hat{R}\,\,\,$  \\ \hline
		40  & Periodic      & 8.88  & 7.85  & 9.75  & 8.77  & \CORR{12.03} & \CORR{11.34} \\
		80  & Quasiperiodic & 11.87 & 8.37  & 12.92 & 10.87 & \CORR{16.76} & \CORR{13.73} \\
		120 & Chaotic       & 13.76 & 9.21  & 16.02 & 12.90 & \CORR{18.70} & \CORR{15.25} \\
	\end{tabular}
	\caption{Mean filter, $E_\mathbf{c}$, velocity, $E_\mathbf{u}$, and temperature, $E_\theta$, errors across dynamical regimes, comparing the baseline isotropic covariances ($Q=\beta I_n$, $R=I_m$) with the data-estimated covariances ($\hat{Q}$, $\hat{R}$). $Pr=10$, $m=48$.}
	\label{tab:cov_comparison}
\end{table}

\section{Conclusions}\label{sec:conclusions}
In our companion paper~\citep{flores2025galerkin}, stable Galerkin ROMs for 2D RB convection were developed using controllability modes of the linearized Boussinesq equations and validated against DNS. The present work has demonstrated a practical application of these ROMs: their use as dynamical models within an EKF for state estimation from sparse measurements.

In \S\ref{sec:coarse_grid}, the filter performance has been evaluated using synthetic measurements extracted from DNS across periodic ($\R = 40$), quasiperiodic ($\R = 80$), and chaotic ($\R = 120$) regimes at $Pr = 10$. With a coarse $4 \times 4$ sensor grid measuring both velocity components and temperature ($m = 48$), the EKF tracks the most energetic modes with high fidelity. Time-averaged reconstruction errors remain below \CORR{$16\%$} for velocity and \CORR{$19\%$} for temperature in all cases. The error increases with the Rayleigh number, consistent with growing dynamical complexity and ROM truncation.

The complementary sensitivity analysis presented in Appendix~\ref{sec:sensitivity} shows that the model-to-measurement noise ratio $\beta = 0.01$ is nearly optimal for the chaotic case and remains a reasonable choice across the other configurations tested. Increasing the observation density yields larger improvements in velocity reconstruction than increasing the ROM dimension, whereas the temperature field is more sensitive to the fidelity of the dynamical model. The best overall performance, with all mean errors below \CORR{$10\%$}, is obtained by combining a 192-mode ROM with $m=192$ observations. \CORR{The robustness of the estimator to instrumental noise has been assessed in Appendix~\ref{sec:noise} by adding white Gaussian noise to the synthetic observations. For noise amplitudes of up to $20\%$ of the RMS value of each measured field ($\mathrm{SNR} \simeq14$\,dB), all mean errors increase by less than five percentage points in the chaotic regime.}

In \S\ref{sec:velocity_only}, the ROM-based EKF has been applied to a hybrid simulation scenario where the state is assimilated from coarse velocity measurements alone. \CORR{Using a $16 \times 16$ grid of velocity probes and no temperature sensors, the filter achieves velocity and temperature errors around $10\%$ and $18\%$, respectively, in the chaotic regime.} These results are encouraging, although real experimental settings involve 3D flows and noisy velocity measurements. Future extensions addressing these challenges may enable the reduction of temperature instrumentation in RB experiments.

In \S\ref{sec:sensor_placement}, a greedy sensor placement algorithm based on the time-averaged Kalman gain norm has been proposed. The algorithm reveals a clear hierarchy in sensor importance: temperature probes are eliminated first, followed by wall-normal velocity sensors, while horizontal velocity measurements are retained until last. 
The greedy strategy is robust across DNS realizations with different initial conditions and substantially outperforms uniform grids of the same observation dimension. The retained sensors tend to cluster in regions of moderate-to-high velocity fluctuations near the walls, indicating that these dynamically active zones are the most informative measurement locations. 

Finally, in \S\ref{sec:estimated_cov} we have relaxed the isotropic, uncorrelated assumption on the noise covariances by estimating $Q$ and $R$ directly from DNS data. Using data-estimated covariances reduces all error metrics across the three dynamical regimes, with filter and velocity errors decreasing by roughly $4.5\%$ and $3\%$ in the chaotic case.

Several directions for future work are envisioned. These include extension to three-dimensional configurations, coupling with state feedback laws for closed-loop control of thermal convection, application to experimental PIV data with noisy measurements, and refinement of the sensor placement strategy using more formal optimization techniques. 
A further direction is the treatment of temporally correlated measurement noise through coloured-noise models.
Moreover, as the implementation of the present EKF is general, it can readily be applied for other flows for which a quadratic Galerkin ROM is available.

\section*{Acknowledgments}
A.V.G.C. acknowledges support from CNPq through grant 314927/2023-9.

{\paragraph*{Data availability statement.} The computational tools used in this study, including Python scripts for modal basis generation and ROM simulation, are made available as an open-source package at \url{https://github.com/kikeflores96/rayleigh_benard_rom.git}. The rest of the data and codes used to generate the results of the present work can be provided upon reasonable request to the authors.}
\appendix

\section{Sensitivity analysis} \label{sec:sensitivity}

\subsection{Influence of noise ratio} 
\begin{figure}[h]
	\centering
	\nolinenumbers
	\begin{subfigure}[b]{0.35\textwidth}
		\centering
		\includegraphics[width=\linewidth]{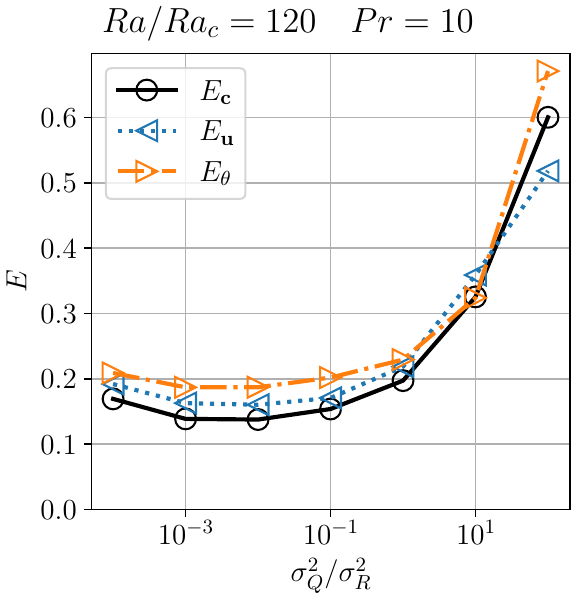}
		\caption{Total errors as a function of $\beta$ in the coarse-grid sensing strategy.}
		\label{fig:QR_ratio_influence}
	\end{subfigure}
	\hfill
	\begin{subfigure}[b]{0.63\textwidth}
		\centering
		\includegraphics[width=\linewidth]{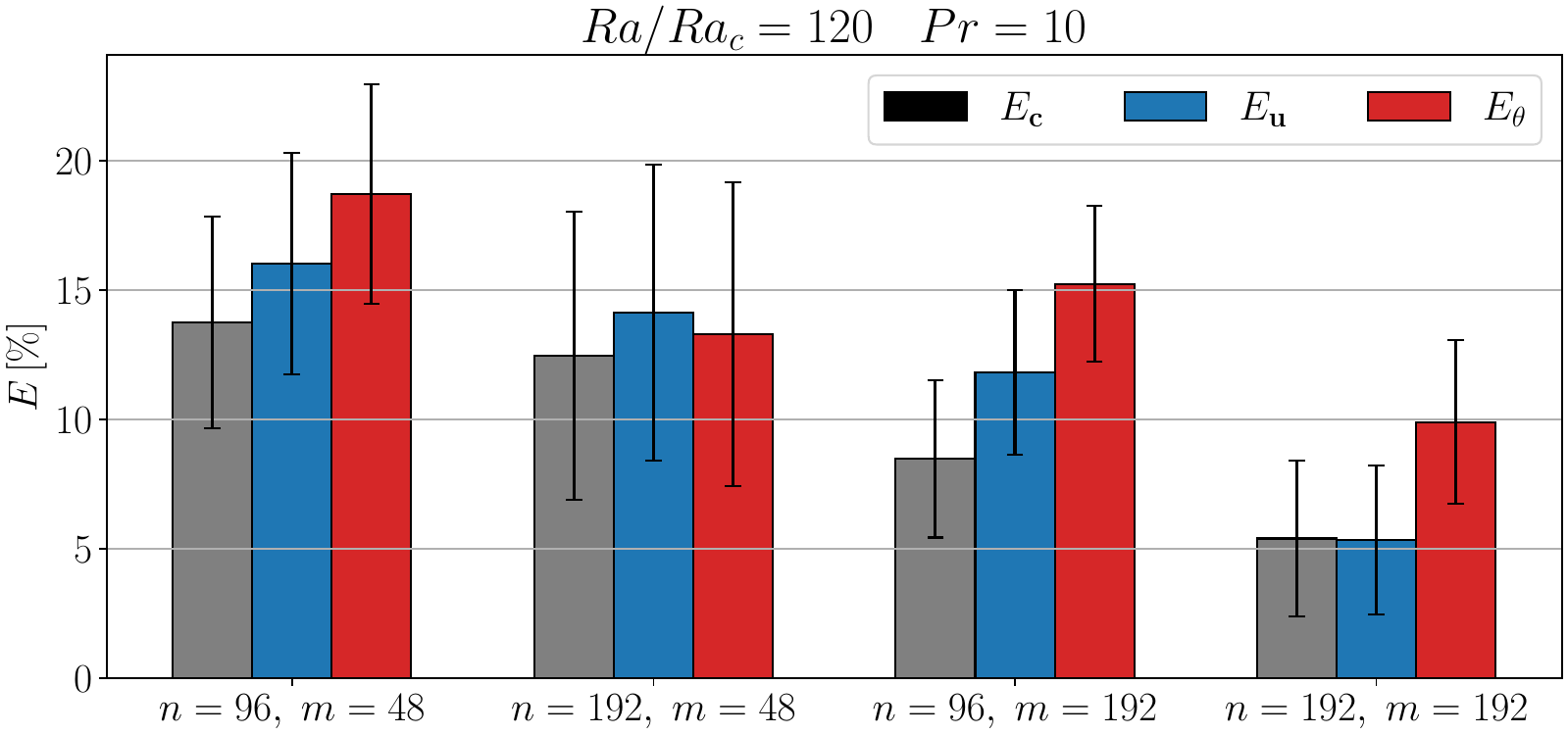}\vspace{2.5em}
		\caption{Error metrics for different ROM dimensions $n$ and observation dimensions $m$.}
		\label{fig:global_comparison}
	\end{subfigure}
	\caption{Sensitivity analysis for the chaotic case $\R=120$, $Pr=10$. (a) Influence of the model-to-measurement noise ratio $\beta$. (b) Influence of the ROM dimension $n$ and the observation dimension $m$. Error bars represent the \CORR{standard deviation of the instantaneous error signals.}}
	\label{fig:sensitivity_combined}
\end{figure}

The performance of the Kalman filter depends on the choice of the noise covariance matrices $Q$ and $R$, which balance the relative confidence placed in the dynamical model and the observations, respectively. This section analyses the sensitivity of the filter to the model-to-measurement noise ratio, $\beta = \sigma_Q^2/\sigma_R^2$ in order to identify an optimal value that is representative across the configurations considered in this study. 
Figure~\ref{fig:QR_ratio_influence} shows the variation of total errors with $\beta$. The analysis has been performed for the chaotic case ($Pr=10$ and $\R=120$) using the coarse-grid strategy presented in \S\ref{sec:coarse_grid}.

The results show that the errors are minimal at approximately $\beta\simeq0.01$, and this value is therefore adopted throughout the paper. The sensitivity of the filter to $\beta$ is asymmetric: errors increase more steeply for $\beta>0.01$, where the filter over-trusts the measurements and the dynamical model is underweighted, than for $\beta<0.01$, where the filter over-trusts the model at the expense of the observations. It is important to note that this trend in the influence of the model-to-measurement noise ratio is approximately preserved across different Rayleigh numbers and observation strategies. While we cannot guarantee that $\beta=0.01$ is strictly optimal in all configurations, it has been observed to consistently outperform both higher and lower values across a wide range of parameter settings, and is therefore retained as a fixed parameter throughout the study.

\subsection{Influence of model and output dimensions} 

This section presents a sensitivity analysis of the Kalman filter to the ROM dimension, $n$, and observation space dimension, $m$. The goal is to assess whether accuracy gains are more effectively achieved by enriching the dynamical model or by increasing the density of the sensor grid. The analysis focuses on the total filter, temperature, and velocity errors, and the observation strategy consists of a decimation of the DNS grid. 

The results are presented for the chaotic case $Pr=10$ and $\R=120$ in Fig.~\ref{fig:global_comparison} using a bar chart. In Fig.~\ref{fig:global_comparison} error bars represent \CORR{the standard deviation of the instantaneous error signals}. Increasing the model dimension from $n=96$ to $n=192$ yields a moderate reduction in all error metrics ($\Delta E_\mathbf{c}=-1.3\%$, $\Delta E_\mathbf{u}=-1.9\%$ and \CORR{$\Delta E_\theta=-5.4\%$}), indicating that the baseline ROM already captures the dominant dynamics and that further enrichment of the modal basis provides diminishing returns. In contrast, increasing the observation dimension from $m=48$ to $m=192$ -- corresponding to a twofold increase in grid resolution in all directions -- substantially improves both the filter and velocity errors (\CORR{$\Delta E_\mathbf{c}=-5.3\%$} and $\Delta E_\mathbf{u}=-4.2\%$). 

\begin{figure}[t]
	\centering
	\nolinenumbers
	\includegraphics[width=\textwidth]{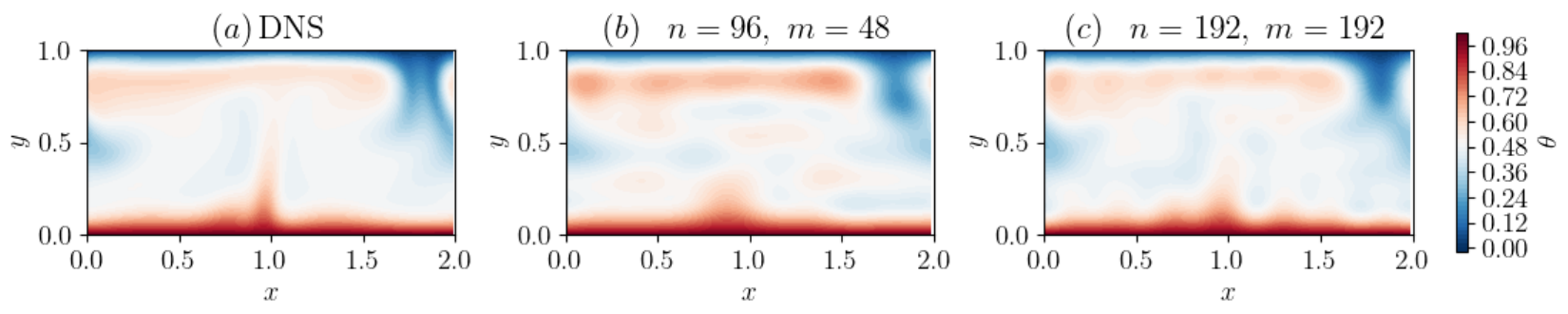}
	\caption{Instantaneous temperature field, $\theta$: (a) DNS, (b) Kalman filter reconstruction with $n=96$ ROM modes and $m=48$ observations, (c) Kalman filter reconstruction with $n=192$ ROM modes and $m=192$ observations. Results are shown for $Ra/Ra_c = 120$ and $Pr = 10$.}
	\label{fig:temperature_comparison}
\end{figure}

However, the reduction in temperature error achieved by refining the observation grid (\CORR{$\Delta E_\theta=-3.5\%$}) is smaller than that obtained by improving the model accuracy (\CORR{$\Delta E_\theta=-5.4\%$}), suggesting that the reconstruction of the thermal field is more sensitive to the fidelity of the dynamical model than to the density of \CORR{measurements}.

This behavior can be understood by considering the structure of the flow at the $Ra$ numbers investigated here: the temperature field consists of thin vertical plumes driven by two counter-rotating vortices. These coherent structures are localized in space and contain relatively high wavenumber components, which require a sufficiently large number of ROM modes to be accurately represented. Therefore, for large modal truncations, the temperature error is partly a consequence of the limited wavenumber content of the ROM basis rather than of insufficient observational coverage. 
In Fig.~\ref{fig:temperature_comparison}, the reconstructed instantaneous temperature field for the two extreme configurations has been represented together with the DNS reference. This illustrates the progressive recovery of the plume structure as both the ROM dimension and number of sensors are increased.
Therefore, for the temperature field, increasing the observation density alone yields lower accuracy gains than improving the dynamical model, since denser observations cannot compensate for modes that are absent from the ROM basis, whereas enriching the latter directly addresses this deficiency.
Finally, a combination of both improvements -- setting $n=m=192$ -- yields the best overall performance, \CORR{reducing the filter and velocity errors to $5\%$ and the temperature error to $10\%$.}  These error levels are remarkably low given the computational efficiency of the ROM-based estimator.

\section{Structure of the estimated noise covariances}\label{sec:cov_structure}
This appendix examines the structure of the data-estimated covariances $\hat{Q}$ and $\hat{R}$ introduced in \S\ref{sec:estimated_cov}, and relates their overall scale to the scalar noise ratio $\beta$ used throughout the study.
Covariance matrices quantify both the magnitude of the error in each component and their correlation.
To isolate the latter, we normalize each matrix into a correlation matrix, $\rho_Q = D^{-1/2}\hat{Q}\,D^{-1/2}$ (and analogously for $\hat{R}$), where $D = \hat{Q}_{ii}$ is the diagonal matrix of variances. This rescaling divides each entry by the standard deviations of the two components involved, setting the diagonal to unity and leaving the off-diagonal entries as correlation coefficients. The off-diagonal structure then reflects only the strength of the cross-correlations, independently of the large magnitude differences between components. 

\begin{figure}[h]
	\centering
	\nolinenumbers
	\includegraphics[width=0.89\linewidth]{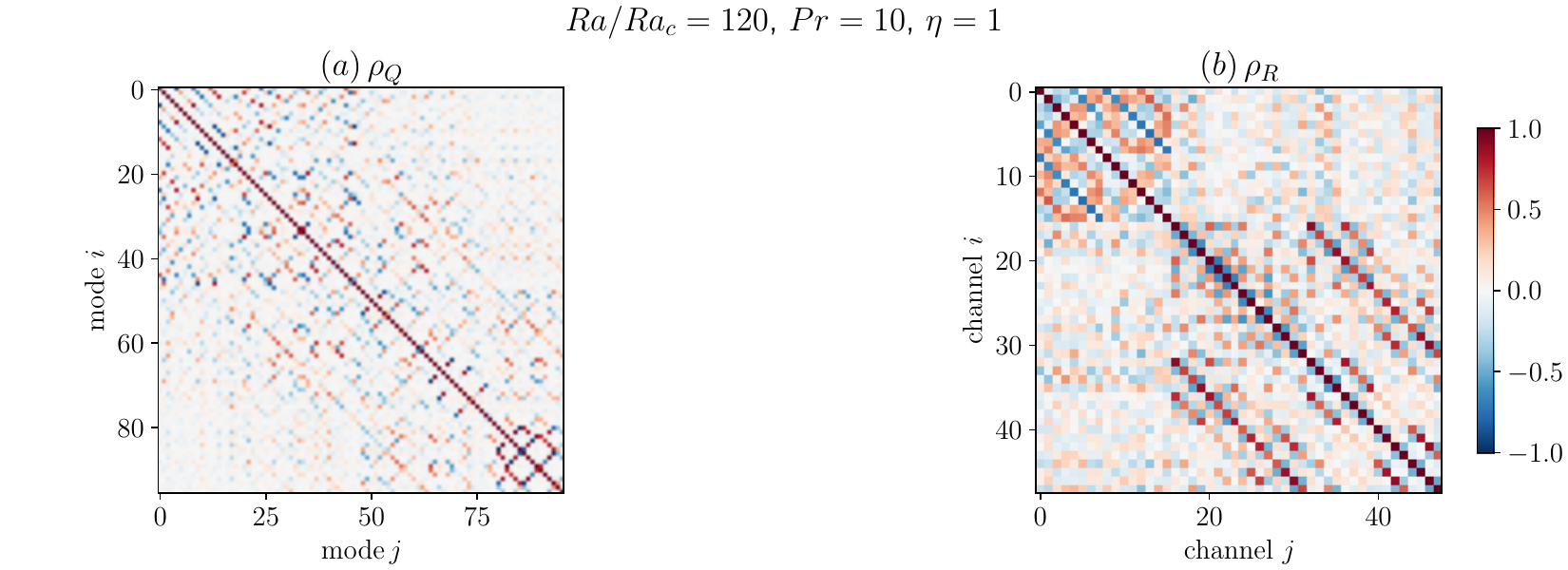}
	\caption{Correlation matrices of the estimated model-error, $\hat{Q}$, and measurement-error, $\hat{R}$, covariances, for the chaotic case $\R = 120$, $Pr = 10$, $\eta = 1$, with the $m = 48$ coarse-grid.}
	\label{fig:cov_correlation}
\end{figure}

Figure~\ref{fig:cov_correlation} shows the two correlation matrices for the chaotic case $\R = 120$, $Pr = 10$, $\eta = 1$, with the $m = 48$ coarse-grid observation strategy. The measurement channels in (b) are ordered as $\left[u,\,v,\,\theta^\prime\right]$ probes. The model-error correlation in Fig.~\ref{fig:cov_correlation}a is dominated by the diagonal but there are multiple non-negligible off-diagonal terms. This indicates that the model errors in different modes are not fully uncorrelated. The measurement-error correlation in Fig.~\ref{fig:cov_correlation}b is far from being diagonal, exhibiting strong cross-correlations between channels. The horizontal velocity probes are highly correlated but there is also a remarkable correlation between some wall-normal velocity and temperature channels. Both matrices therefore deviate from the isotropic, uncorrelated assumption of the baseline filter, which explains the performance gains observed in \S\ref{sec:estimated_cov}.
It is worth relating these estimated matrices to the scalar ratio $\beta$ used elsewhere in the study. It is possible to obtain a data-implied noise ratio, $\hat{\beta}$, from the two covariances using the matrix Frobenius norm normalized by the square root of its dimension:
\begin{equation}\label{eq:beta_hat}
	\hat{\beta} = \frac{\|\hat{Q}\|\,/\,\sqrt{n}}{\|\hat{R}\|\,/\,\sqrt{m}} = \frac{\|\hat{Q}\|}{\|\hat{R}\|}\sqrt{\frac{m}{n}}.
\end{equation}
This definition recovers the noise ratio value in the isotropic uncorrelated limit. Equation~\eqref{eq:beta_hat} applied to the estimated covariance matrices yields $\hat{\beta} \approx 0.00945$, in close agreement with the value $\beta = 0.01$ identified by the sensitivity analysis of Appendix~\ref{sec:sensitivity}. This consistency indicates that the hand-tuned ratio used throughout the paper is close to the balance implied by the data.

\CORR{
\section{Influence of measurement noise}\label{sec:noise}

Throughout this work, the synthetic observations have been extracted directly from the DNS fields. Therefore, for an exact estimate of the projected state, $\boldsymbol{c}_k$, the discrepancy between the measurements and the filter output is given by the off-projection error, $\boldsymbol{y}_k - H\boldsymbol{c}_k$. However, real sensors are also subject to instrumental noise.  
In this appendix, we introduce controlled-amplitude noise in the observations to assess the robustness of the EKF in a more realistic scenario. The noisy filter measurements, $\tilde{\boldsymbol{y}}_k$, are given by:
\begin{equation}\label{eq:noisy_obs}
	\tilde{\boldsymbol{y}}_k = \boldsymbol{y}_k + \boldsymbol{\xi}_k, \qquad \boldsymbol{\xi}_k \sim \mathrm{N}(\boldsymbol{0},\,\Sigma),
\end{equation}
where $\boldsymbol{y}_k$ are the noise-free DNS observations defined in eq.~\eqref{eq:meas_vect} and $\boldsymbol{\xi}_k$ is an additive Gaussian noise, independent across the different measurement channels and white in time. Unlike a multiplicative noise model, which scales with the instantaneous value of the signal and vanishes wherever the signal is small, the additive model introduces a constant noise floor, independent of the measured value. The noise covariance is diagonal, with one standard deviation per measured variable:
\begin{equation}\label{eq:noise_cov}
	\Sigma = \mathrm{diag}\left(\sigma_u^2 I_{n_u},\; \sigma_v^2 I_{n_v},\; \sigma_{\theta'}^2 I_{n_\theta}\right).
\end{equation}
The noise amplitude of each variable is prescribed as a fraction, $\epsilon$, of the root-mean-square (RMS) value of the corresponding field over the fluid domain and the assimilation window:
\begin{equation}\label{eq:noise_amplitude}
	\sigma_q = \epsilon\, q_\mathrm{rms}, \qquad
	q_\mathrm{rms} = \left(\frac{1}{T}\int_0^T \frac{1}{L_xL_y}\int_\Omega q^2\,\diff\Omega\,\diff t\right)^{1/2}, \qquad
	q \in \{u,\,v,\,\theta'\}.
\end{equation}
Hereafter, $\epsilon$ is referred to as the noise level. This definition links the noise level to the signal-to-noise ratio (SNR) of each measured variable. Defining the SNR as the ratio between the mean signal power and the noise variance yields:
\begin{equation}\label{eq:SNR}
	\mathrm{SNR} = \frac{q_\mathrm{rms}^2}{\sigma_q^2} = \frac{1}{\epsilon^2},
	\qquad
	\mathrm{SNR}_\mathrm{dB} = 10\log_{10}\mathrm{SNR} = -20\log_{10}\epsilon.
\end{equation}
With this definition, the SNR is identical for the three measured variables. Note that eq.~\eqref{eq:SNR} defines the SNR with respect to the mean power of each field over the whole domain. Consequently, the local SNR of an individual probe is lower in regions of weak signal, e.g.\ close to the walls, and higher in dynamically active regions. In the following, five noise levels are considered: $\epsilon = 1\%$, $2\%$, $5\%$, $10\%$ and $20\%$, corresponding to $\mathrm{SNR}_\mathrm{dB} = 40$, $34$, $26$, $20$ and $14$\,dB, respectively. The filter parameters are kept identical to those of \S\ref{sec:coarse_grid}. In particular, the noise covariances are maintained at $Q = \beta I_n$ and $R = I_m$ with $\beta = 0.01$, so that the filter is not informed of the added noise. The analysis is performed for the chaotic case ($\R = 120$, $Pr = 10$) with random seed $\eta = 1$, using the $m = 48$ coarse-grid observation strategy.
\begin{figure}[h]
	\centering
	\nolinenumbers
	\includegraphics[width=0.5\linewidth]{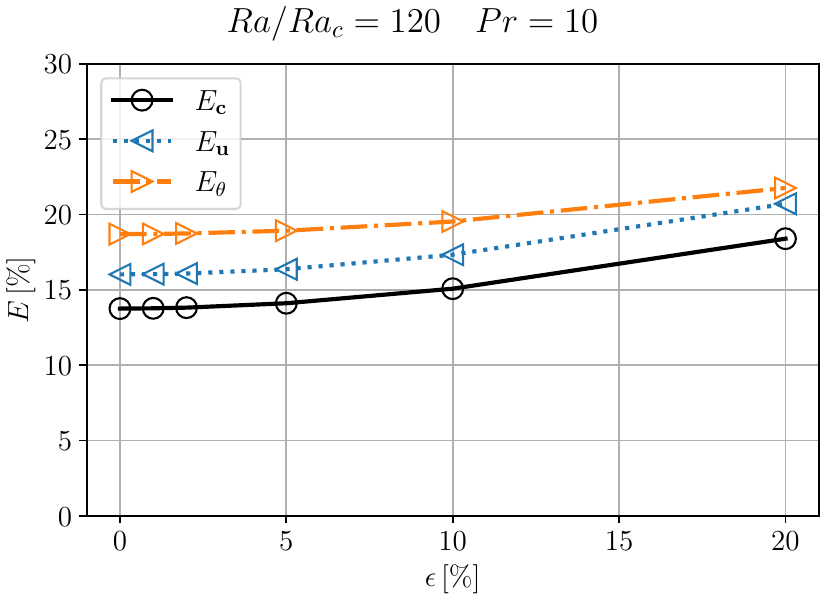}
	\caption{Total filter, $E_\mathbf{c}$, velocity, $E_\mathbf{u}$, and temperature, $E_\theta$, errors as a function of the measurement noise level $\epsilon$ for the chaotic case ($\R=120$, $Pr=10$, $\eta = 1$), with the $m=48$ coarse-grid observation strategy and $\beta = 0.01$.}
	\label{fig:noise_influence}
\end{figure}

Figure~\ref{fig:noise_influence} shows the total errors as a function of the noise level. The noise-free case, $\epsilon = 0$, recovers the values reported in Tab.~\ref{tab:errors}. The three error metrics increase with the noise level: from $\epsilon=0$ to $\epsilon=20\%$, the filter error grows from $13.8\%$ to $18.4\%$, the velocity error from $16.0\%$ to $20.7\%$ and the temperature error from $18.7\%$ to $21.8\%$. These results indicate that the present estimator is robust to moderate levels of measurement noise, which is an encouraging result for its application to experimental data.
}

\bibliographystyle{unsrtnatabbrv}
\bibliography{references}
\end{document}